\documentclass[12pt,nofootinbib,superscriptaddress]{revtex4}

\usepackage{amssymb,color,paralist,amsmath,epsfig}
\usepackage{graphicx}
\usepackage{amsfonts}
\usepackage{nicefrac,esint}
\usepackage{verbatim}
\usepackage{float}
\usepackage{comment}
\usepackage{relsize}
\usepackage{mathrsfs}

\usepackage[colorlinks,citecolor=blue,linktoc=all,linkcolor=cyan]{hyperref}

\newcommand{\beq}{\begin{equation}}
\newcommand{\eeq}{\end{equation}}

\newcommand{\ber}{\begin{eqnarray}}
\newcommand{\eer}{\end{eqnarray}}

\begin{document}

\rightline{FERMILAB-PUB-20-418-T}

\title{Axial and pseudoscalar form factors from charged current quasielastic neutrino-nucleon scattering}

\author{Oleksandr Tomalak}
\affiliation{Department of Physics and Astronomy, University of Kentucky, Lexington, Kentucky 40506, USA}
\affiliation{Fermilab, Batavia, Illinois 60510, USA}

\date{\today}

\begin{abstract}
We study the scattering of neutrinos on polarized and unpolarized free nucleons, and also the polarization of recoil particles in these scatters. In contrast to electromagnetic processes, the parity-violating weak interaction gives rise to large spin asymmetries at leading order. Future polarization measurements could provide independent access to the proton axial structure and allow the first extraction of the pseudoscalar form factor from neutrino data without the conventional partially conserved axial current (PCAC) ansatz and assumptions about the pion-pole dominance. The pseudoscalar form factor can be accessed with precise measurements with muon (anti)neutrinos of a few hundreds $\mathrm{MeV}$ of energy or with tau (anti)neutrinos. The axial form factor can be extracted from scattering measurements using accelerator neutrinos of all energies.
\end{abstract}
\maketitle

\tableofcontents

\section{Introduction}
\label{sec1}

Neutrino physics is entering a precision era driven by new experiments and modern detector technology. This requires an improved theoretical and phenomenological description of neutrino interactions. An ambitious goal of percent level measurements calls for precise inputs from nuclear and hadronic physics~\cite{Formaggio:2013kya,Mosel:2016cwa,Alvarez-Ruso:2017oui}. To describe elementary neutrino-nucleon charged current quasielastic (CCQE) interactions inside the nucleus, four nucleon form factors have to be precisely known. These are the isovector electric and magnetic form factors, the axial form factor, and the pseudoscalar form factor. The former pair can be precisely measured in electron scattering. References~\cite{Bernauer:2010wm,Bernauer:2013tpr,Xiong:2019umf,Punjabi:2015bba,Ganichot:1972mb,Bosted:1989hy} provide data at low momentum transfer corresponding to enhanced event rates in neutrino experiments. The axial and pseudoscalar form factors require weak probes with neutrinos~\cite{Rock:1991jy,Mann:1973pr,Barish:1977qk,Miller:1982qi,Baker:1981su,Belikov:1983kg,Bernard:2001rs,Kitagaki:1983px,Kitagaki:1990vs,Milbrath:1997de,Wright:1998gi,Gorringe:2002xx,Winter:2011yp,Bhattacharya:2011ah,Andreev:2012fj,Andreev:2015evt,Hill:2017wgb} often accompanied with nuclear physics effects~\cite{Lyubushkin:2008pe,Benhar:2009wi,Martini:2009uj,Nieves:2011yp,Nieves:2011sc,Anderson:2011ce,Benhar:2014qaa,Wolcott:2015yja,Devan:2015uak,Gallmeister:2016dnq,HurtadoAnampa:2016uqe,Grover:2018out,Rocco:2018mwt,Nikolakopoulos:2019qcr,Gonzalez-Jimenez:2019qhq,Lovato:2020kba,Abe:2020jbf,King:2020cvj} or measurements of pion electroproduction~\cite{Choi:1993vt,Bernard:1994pk,Blomqvist:1996tx,Liesenfeld:1999mv,Kamalov:2001qg,Gran:2006jn,Friscic:2015tga}. The axial form factor which is known to $10-20~\%$ is the main source of error in microscopic description of neutrino interactions at the nucleon level. Improved measurements of the axial form factor,  ideally with independent systematic uncertainties, and model-independent extractions of the pseudoscalar form factor are important for understanding nucleon dynamics at momentum transfers $Q^2\lesssim 1-3~\mathrm{GeV}^2$ and are essential for modeling of neutrino interactions at DUNE~\cite{Alion:2016uaj,Abi:2020evt}, Hyper-K~\cite{Hyper-Kamiokande:2016dsw}, and ESS$\nu$SB~\cite{Baussan:2013zcy}.

Besides unpolarized cross section measurements, one can perform experiments with polarized particles~\cite{Dombey:1969wk,Akhiezer:1974em} and access form factors in a complementary way as was successfully realized in electron-proton scattering a few decades ago~\cite{Perdrisat:2006hj,Jones:1999rz,Gayou:2001qd,Punjabi:2005wq,Puckett:2010ac,Ron:2011rd,Zhan:2011ji}. After pioneering studies of polarization observables in neutrino physics~\cite{Lee:1962jm,Adler:1963,Florescu:1968zz,Pais:1971er,Cheng:1971mx,Tarrach:1974da,Oliver:1974de,Kim:1978fx,Ridener:1984tb,Ridener:1986ey}, a few groups have recently revisited polarization effects in (anti)neutrino-nucleon charged current quasielastic scattering~\cite{Bilenky:2013fra,Bilenky:2013iua,Graczyk:2019xwg,Graczyk:2019opm}.\footnote{In the following, charged current quasielastic scattering refers to processes on free nucleons $\nu_\ell n \to \ell^- p$ and $\bar{\nu}_\ell p \to \ell^+ n$. Further nuclear physics effects are beyond the scope of this work.} Expressions for all possible single-, double- and triple-spin asymmetries in scattering on free nucleons are collected in~\cite{Graczyk:2019xwg}. The contributions of second-class currents to polarization observables are considered in~\cite{Fatima:2018tzs}. Polarization effects in inverse reactions $e p \to \nu n$ are described in~\cite{Fatima:2018gjy}. The discovery of the tau neutrino~\cite{Kodama:2000mp} and subsequent experiments~\cite{Kodama:2007aa,Agafonova:2014bcr,Bonivento:2017acp} have motivated studies of CCQE observables with polarized recoil tau leptons (or just taus)~\cite{Hagiwara:2003di,Hagiwara:2004gs,Graczyk:2004vg,Graczyk:2004uy,Bourrely:2004iy,Kuzmin:2004ke,Kuzmin:2004yb,Aoki:2005wb,Aoki:2005kc,Sobczyk:2019urm,Fatima:2020pvv}. Induced nucleon polarization in (anti)neutrino-nucleus neutral current scattering are described in~\cite{Jachowicz:2004we,Jachowicz:2005np,Lava:2006fc,Meucci:2008zz}.

In many conventional treatments of neutrino-nucleon interactions, the pseudoscalar form factor is related to the axial one using the partially conserved axial current (PCAC) ansatz and assuming the pion-pole dominance, which is only expected to be a valid approximation at low momentum transfers~\cite{LlewellynSmith:1971uhs,Bernard:1998gv,Fuchs:2002zz,Kaiser:2003dr,Schindler:2006it,Lutz:2020dfi,Chen:2020wuq}. Recent advances in lattice QCD have provided us with {\it{ab initio}} results both for the axial and for the pseudoscalar form factors~\cite{Yamazaki:2007mk,Alexandrou:2009vqd,Bratt:2010jn,Alexandrou:2010hf,vonHippel:2016nfg,Capitani:2017qpc,Green:2017keo,Alexandrou:2017hac,Alexandrou:2018pln,Bali:2019yiy,Hasan:2019noy}. Though initially there was a strong disagreement with the PCAC ansatz in the assumption of pion-pole dominance~\cite{Jang:2016kch,Rajan:2017lxk}, this problem seems to be resolved with recent calculations satisfying the PCAC ansatz within the statistical errors of simulations~\cite{Bali:2018qus,Jang:2019vkm,Bali:2019yiy,Bar:2019igf,Bar:2019gfx,Bhattacharya:2020zjf,Park:2020axe,Jang:2020ygs} even at $Q^2\sim 1~\mathrm{GeV}^2$. However, the other recent lattice calculation~\cite{Alexandrou:2020okk} at the physical pion mass has indicated on the violation of the PCAC at low momentum transfer. In experiment, the value of $F_P(Q^2)$ at $Q^2 \sim 0.88 m_\mu^2$, where  $m_\mu$ is the mass of the muon, can be extracted from measurements of  muon capture on the proton~\cite{Wright:1998gi,Winter:2011yp,Andreev:2012fj,Andreev:2015evt}, reviews are available in~\cite{Gorringe:2002xx,Hill:2017wgb}. At other values of momentum transfer, the pseudoscalar form factor was extracted only once from the pion electroproduction cross section data~\cite{Choi:1993vt,Bernard:1994pk}.

Novel extractions of the axial and the pseudoscalar form factors from neutrino experiments with polarized particles were recently proposed in the Snowmass 2021 Letter of Intent ``Neutrino Scattering Measurements on Hydrogen and Deuterium"~\cite{Snowmass2021LoInuHD}. Experiments on hydrogen and deuterium targets, when nuclear physics effects are absent or can be taken under control, provide relatively clean and unambiguous probes of nucleon axial structure over a wide range of $Q^2$ while still allowing sizable event rates. 

In this work, we study the sensitivity of single-spin asymmetries in (anti)neutrino charged current quasielastic scattering on free nucleons to the axial and the pseudoscalar form factors. We determine neutrino beam energies suitable for the simultaneous extraction of both form factors in a single experiment and identify single-spin asymmetries sensitive to the axial contributions at GeV energies. In what follows we calculate spin-polarized observables for target nucleon's, recoiling nucleon's, and recoiling lepton's polarizations. We describe the most promising channels for extracting the pseudoscalar and axial nucleon form factors. 

This paper is organized as follows. In Sec.~\ref{sec2}, we express the unpolarized cross section and single-spin asymmetries in terms of nucleon form factors in a simple and convenient way. In Sec.~\ref{sec3}, we study single-spin asymmetries and unpolarized cross section for muon neutrino beam of hundred MeV energies. We concentrate on prospects for extraction of the axial form factor with polarization observables at GeV energies in the following Sec.~\ref{sec4}. In Sec.~\ref{sec5}, we investigate the unpolarized cross section and all single-spin asymmetries with a beam of tau neutrinos. Sec.~\ref{sec6} provides conclusions. For the convenience of the reader, we provide Figs.~\ref{fig:Tt_asymmetries}$-$\ref{fig:Ll_asymmetries2} with polarization observables and unpolarized cross sections.

\section{Nucleon form factors in unpolarized cross section and polarization observables}
\label{sec2}

We present a relation between microscopic physics and experimental observables in this Section. First, we define nucleon form factors as matrix elements of quark currents. Based on this definition, we express the unpolarized cross section and single-spin asymmetries in terms of nucleon form factors. In this paper, we generalize well-known for the unpolarized cross section structure-dependent parameters $A, B$, and $C$ to single-spin asymmetries.

For neutrinos with energies typical in accelerator-based experiments, corresponding to kinematics much below the electroweak scale, charged current (anti)neutrino-quark scattering is described by the four-fermion interaction:
\ber
    {\cal L}_{\rm eff} =
  - \sum_{q \ne q^\prime} \left( c_{q q'} 
  \bar{\ell} \gamma^\mu \mathrm{P}_\mathrm{L} \nu_{\ell}
  \, \bar{q} \gamma_\mu \mathrm{P}_\mathrm{L}  q^\prime + \mathrm{h.c.} \right)\, ,  \label{effective_Lagrangian}
\eer
where $\mathrm{P}_\mathrm{L}$ is the projection operator on the left-handed chiral states. At leading order, the Wilson coefficients $c_{q q'}$ are given by $2 \sqrt{2} \mathrm{G}_\mathrm{F} V_{q q^\prime}$, where $\mathrm{G}_\mathrm{F}$ is the Fermi coupling constant and $V_{q q^\prime}$ is the Cabibbo-Kobayashi-Maskawa (CKM) matrix element. A more precise determination of the Wilson coefficients is given in~\cite{Hill:2019xqk}. It is beyond the level of accuracy considered in this paper and we exploit the tree-level values quoted above.

We consider neutrino-neutron and antineutrino-proton CCQE scattering:
\ber
\nu_\ell \left( k \right) n \left( p \right) &\to& \ell^- \left( k' \right) p \left( p' \right), \\
\bar{\nu}_\ell \left( k \right) p \left( p \right) &\to& \ell^+ \left( k' \right) n \left( p' \right),
\eer
with four-momenta of incoming and outgoing particles $k,p$ and $k',p'$, respectively. The matrix element of the quark current $\bar{u} \gamma_\mu \mathrm{P}_L d$ inside the nucleon in neutrino-neutron CCQE $\nu_\ell n \to \ell^- p $, can be expressed in terms of Sachs electric, $G^V_E$, and magnetic, $G^V_M$, isovector, axial, $F_A$, and pseudoscalar, $F_P$, form factors as~\cite{LlewellynSmith:1971uhs}\footnote{Our definition of form factors implies $F_A(0) < 0$.}
\ber
\Gamma_\mu (Q^2) =  \langle p(p^\prime) | \bar{u} \gamma_\mu \mathrm{P}_L d | n(p) \rangle  &=& \frac12 \bar{p}\! \left[ 
\gamma_\mu G_M^V(Q^2) - \frac{p_\mu + p'_\mu}{2M} \frac{G_M^V(Q^2)-G_E^V(Q^2)}{1+\tau} \right]\! \nonumber \\
&+&\frac12 \bar{p}\! \left[  \gamma_\mu \gamma_5 F_A(Q^2) + \frac{q_\mu}{M} \gamma_5 F_P(Q^2) \right]\! n \,,\label{eq:FFs}
\eer
with $q = p' - p$, $Q^2 = - \left( p - p'\right)^2$ and $\tau=Q^2/(4M^2)$. In the limit of isospin symmetry, when the mass, $M$, for both nucleons is approximately the same, both the electric and magnetic isovector form factors are given by the difference of proton and neutron form factors, i.e., $G_{E,M}^V = G_{E,M}^p-G_{E,M}^n$.\footnote{Isospin-breaking effects are below the level of precision in this work.} In this paper, we assume $CP$ invariance and work at leading order in the QED coupling constant, when all form factors in Eq.~(\ref{eq:FFs}) are real functions. Antineutrino-proton CCQE scattering $\bar{\nu}_\ell p  \to \ell^+  n $ is described by the conjugated current.

We consider a few experimental observables in the following. The unpolarized (anti)neutrino-nucleon scattering cross section is conveniently expressed in terms of the structure-dependent $A,B,C$ parameters~\cite{LlewellynSmith:1971uhs,Formaggio:2013kya}
\begin{align}
\frac{d\sigma}{dQ^2} (Q^2, E_\nu) = \frac{c^2_{q q'}}{16\pi} \frac{M^2}{E_\nu^2}
\left[ \left( \tau + r^2 \right)A(Q^2) -  \nu B(Q^2) + \frac{\nu^2}{1+\tau} C(Q^2)  \right] \,,
\label{eq:xsection_CCQE}
\end{align}
where $r = m_\ell/(2 M)$ with the lepton mass $m_\ell$, incoming neutrino energy $E_\nu$ and variable $\nu = E_\nu/M -  \tau - r^2$. The structure-dependent factors $A,~B$, and $C$ are given by
\begin{align}\label{eq:ABC}
   A &= \tau \left( G^V_M \right)^2 - \left( G^V_E \right)^2 + (1+ \tau) F_A^2 - r^2 \left(  \left( G^V_M \right)^2 + F_A^2 - 4 \tau F_P^2+ 4 F_A F_P   \right) \,, \\
 B &=  4 \eta \tau F_A G^V_M \,, \\
 C &=  \tau  \left( G^V_M \right)^2 + \left( G^V_E \right)^2 +  (1+ \tau) F_A^2 \,,
\end{align}
where $\eta=+1$ corresponds to neutrino scattering $\nu_\ell n \to \ell^- p$ and $\eta=-1$ corresponds to antineutrino scattering $\bar{\nu}_\ell p \to \ell^+ n$. The momentum transfer increases from forward to backward directions with corresponding values $Q^2_\mathrm{-} $ to $Q^2_\mathrm{+}$, respectively,
\ber
Q^2_\pm &=& \frac{2 M E^2_\nu}{M+2 E_\nu} - 4M^2 \frac{M+E_\nu}{M+2 E_\nu} r^2 \pm \frac{4 M^2 E_\nu}{M+2 E_\nu} \sqrt{\left(  \frac{E_\nu}{2M}- r^2 \right)^2- r^2}.
\eer
The contribution of the pseudoscalar form factor $F_P$ to the unpolarized cross section is suppressed by the lepton mass. At energies of accelerator neutrinos, it is below errors associated with the axial form factor $F_A$.

Besides the unpolarized cross section, various spin-dependent observables can be accessed experimentally. The simplest ones are spin asymmetries which are the main subject of this paper. Target, $\mathrm{T}$, recoil, $\mathrm{R}$, and lepton, $\mathrm{L}$, single-spin asymmetries are defined from the difference of cross section $\sigma \left( \vec{S} \right)$ with a fixed spin direction $\vec{S}$ of one incoming or outgoing particle and cross section $\sigma \left( -\vec{S} \right)$ with the spin in the opposite direction as
\ber \label{eq:asymmetry}
\mathrm{T}, \mathrm{R},\mathrm{L} = \frac{\mathrm{d} \sigma \left( \vec{S}^{\mathrm{T}, \mathrm{R},\mathrm{L}} \right) - \mathrm{d} \sigma \left( - \vec{S}^{\mathrm{T}, \mathrm{R},\mathrm{L}}  \right)}{\mathrm{d} \sigma \left(  \vec{S}^{\mathrm{T}, \mathrm{R},\mathrm{L} } \right) + \mathrm{d} \sigma \left( -\vec{S}^{\mathrm{T}, \mathrm{R},\mathrm{L}}  \right)}. \label{eq:target_asymmetry}
\eer
At leading order in QED when all form factors in Eq.~(\ref{eq:FFs}) are real functions, single-spin asymmetries can be described by two independent spin components in the scattering plane with the spin direction parallel or perpendicular to outgoing reference particle or to the beam direction. Asymmetries are conveniently expressed in terms of new structure-dependent functions that depend on the particle whose spin we are considering:
\ber
\mathrm{T}, \mathrm{R},\mathrm{L} = \frac{\left( \tau + r^2 \right)A^{\mathrm{T}, \mathrm{R},\mathrm{L}}(Q^2) -  \nu B^{\mathrm{T}, \mathrm{R},\mathrm{L}}(Q^2) + \frac{\nu^2}{1+\tau} C^{\mathrm{T}, \mathrm{R},\mathrm{L}}(Q^2)}{\left( \tau + r^2 \right)A(Q^2) -  \nu B(Q^2) + \frac{\nu^2}{1+\tau} C(Q^2)}. \label{eq:asymmetryABC}
\eer

For (anti)neutrino scattering on the polarized nucleon target with the spin four-vector $S$, the asymmetry $\mathrm{T}$ is determined by the following structure-dependent factors $A^\mathrm{T},~B^\mathrm{T}$, and $C^\mathrm{T}$:\footnote{To simplify our expressions, we use an unconventional normalization for the spin four-vector: $S^2 = - 1/M^2$.}
\begin{align} \label{eq:ABC_recoil1}
 A^\mathrm{T} &=G^V_M   \left(   F_A- \eta G^V_E \right) \left( p' \cdot S \right) - 2 \eta G^V_M G^V_E \left( k' \cdot S \right) \nonumber \\
 &+ 2 r^2 G^V_M \left(  \frac{ \eta G^V_E- F_A + 2 \tau F_P }{\tau + r^2} \left( k\cdot S \right) - F_P \left( p' \cdot S \right) \right)  \,, \\
 B^\mathrm{T} &= \left( \eta F_A^2 -  F_A G^V_E +\eta \tau	G^V_M \frac{  G^V_M - G^V_E }{1+\tau} \right)  \left( p' \cdot S \right) - 2  F_A  G^V_E \left( k' \cdot S \right) \nonumber \\
 & - r^2  \left( F_A\frac{  G^V_M - G^V_E }{1+\tau} - 2 F_P \frac{G^V_E + \tau G^V_M}{1+\tau} \right) \left( p' \cdot S \right)  \,, \\
 C^\mathrm{T} &=   F_A \left(  G^V_M - G^V_E \right) \left( p' \cdot S \right)  \,, \label{eq:ABC_recoil3}
\end{align}
where $\eta=+1$ corresponds to neutrino scattering $\nu_\ell n \to \ell^- p$ and $\eta=-1$ corresponds to antineutrino scattering $\bar{\nu}_\ell p \to \ell^+ n$. It  is worthwhile highlighting the special cases. To evaluate $\mathrm{T}_\mathrm{t}$, the asymmetry in which the target polarization is transverse to the beam direction with the spin vector in the scattering plane, we substitute $\left( p' \cdot S \right) = - \left( k' \cdot S \right) = \frac{2M}{E_\nu} \sqrt{\tau \nu^2 -  (1+\tau)  (\tau + r^2)^2 }$ in Eqs.~(\ref{eq:ABC_recoil1})$-$(\ref{eq:ABC_recoil3}) above. To evaluate $\mathrm{T}_\mathrm{l}$, the asymmetry in which the target polarization is along the beam direction, we substitute $\left( p' \cdot S \right) = - 2 \left( \tau + \frac{M}{E_\nu} \left( \tau + r^2 \right)\right)$ and $\left( k' \cdot S \right) = - \left( p' \cdot S \right) - \frac{E_\nu}{M}$ in Eqs.~(\ref{eq:ABC_recoil1})$-$(\ref{eq:ABC_recoil3}) above. The transverse target single-spin asymmetry $\mathrm{T_t}$ vanishes at forward and backward angles. The longitudinal single-spin asymmetry is positive at forward scattering when the momentum transfer is $Q^2_\mathrm{-}$. Up to lepton-mass-suppressed terms, the asymmetry $\mathrm{T_l}$ reaches maximum by an absolute value at backward angles when the momentum transfer is $Q^2_\mathrm{+}$. For these kinematic boundaries, the longitudinal target single-spin asymmetry is given by
\ber
\mathrm{T}_\mathrm{l} \left( Q^2_\mathrm{-} \right) = - \frac{2 G^V_E F_A}{  \left(G^V_E\right)^2 + F_A^2} + \mathrm{O} \left( m_\ell^2 \right), \qquad \mathrm{T}_\mathrm{l} \left( Q^2_\mathrm{+} \right) = \eta + \mathrm{O} \left( m_\ell^2 \right).
\eer

For (anti)neutrino scattering with measurements of the recoil nucleon spin $S$, the asymmetry $\mathrm{R}$ is determined by the following structure-dependent factors $A^\mathrm{R},~B^\mathrm{R}$, and $C^\mathrm{R}$:
\begin{align} \label{eq:ABC_target1}
 A^\mathrm{R} &=G^V_M   \left(  F_A - \eta G^V_E   \right) \left( p \cdot S \right) - 2 \eta G^V_M G^V_E \left( k \cdot S \right) \nonumber \\
 &+ 2 r^2 G^V_M \left(  \frac{  \eta G^V_E + F_A  - 2 \tau F_P }{\tau + r^2} \left( k \cdot S \right) - F_P \left( p \cdot S \right) \right) \,, \\
 B^\mathrm{R} &= \left( \eta F_A^2 -  F_A G^V_E +\eta \tau	G^V_M \frac{  G^V_M - G^V_E }{1+\tau} \right)  \left( p \cdot S \right) - 2 F_A  G^V_E  \left( k \cdot S \right) \nonumber \\
 &+ r^2  \left( F_A\frac{  G^V_M - G^V_E }{1+\tau} - 2 F_P \frac{G^V_E + \tau G^V_M}{1+\tau} \right) \left( p \cdot S \right)  \,, \\
 C^\mathrm{R} &=   F_A \left(  G^V_M - G^V_E \right) \left( p \cdot S \right)  \,. \label{eq:ABC_target3}
\end{align}
To evaluate $\mathrm{R}_\mathrm{t}$, the recoil nucleon spin asymmetry with the spin vector in the scattering plane and perpendicular to the recoiling nucleon's momentum, we substitute $\left( p \cdot S \right) = 0 $ and $\left( k \cdot S \right) = -    \sqrt{ \tau \nu^2 -  (1+\tau)  (\tau + r^2)^2 }/\sqrt{\tau (1+\tau)}$ in Eqs.~(\ref{eq:ABC_target1})$-$(\ref{eq:ABC_target3}) above. To evaluate $\mathrm{R}_\mathrm{l}$, the recoil nucleon spin asymmetry with the spin vector in the scattering plane and parallel to the recoiling nucleon's momentum, we substitute $\left( p \cdot S \right) = 2  \sqrt{\tau (1+\tau)} $ and $\left( k \cdot S \right) =   \left( \tau \nu -  (1+\tau)  (\tau + r^2) \right)/\sqrt{\tau (1+\tau)} $ in Eqs.~(\ref{eq:ABC_target1})$-$(\ref{eq:ABC_target3}) above. The transverse recoil single-spin asymmetry $\mathrm{R_t}$ vanishes at forward and backward angles. The longitudinal single-spin asymmetry is positive at forward scattering when the momentum transfer is $Q^2_\mathrm{-}$. Up to lepton-mass-suppressed terms, the asymmetry $\mathrm{R_l}$ reaches maximum by an absolute value at backward angles when the momentum transfer is $Q^2_\mathrm{+}$. For these kinematic boundaries, the longitudinal target single-spin asymmetry is given by
\ber
\mathrm{R}_\mathrm{l} \left( Q^2_\mathrm{-} \right) = - \frac{2 G^V_E F_A}{  \left(G^V_E\right)^2 + F_A^2}  + \mathrm{O} \left( m_\ell^2 \right), \qquad \mathrm{R}_\mathrm{l} \left( Q^2_\mathrm{+} \right) = - \eta + \mathrm{O} \left( m_\ell^2 \right).
\eer

For (anti)neutrino scattering with measurements of the recoil lepton spin $S$, the asymmetry $\mathrm{L}$ is determined by the following structure-dependent factors $A^\mathrm{L}, B^\mathrm{L}$, and $C^\mathrm{L}$:
\begin{align} \label{eq:ABC_lepton1}
   \left( \tau + r^2 \right) A^\mathrm{L}  &= - \eta A \left( k \cdot rS \right) + 2 \left( \tau + r^2 \right) F_A G^V_M \left( k +2 p\cdot rS \right) \nonumber \\
   &- 2 \eta r^2  \left(  \left( G^V_M \right)^2 + F_A^2 - 4 \tau F_P^2+ 4 F_A F_P   \right)  \left( k \cdot rS \right)\,, \\
   B^\mathrm{L}  &=  -2 F_A G^V_M \left( k \cdot rS \right) + \frac{\eta C}{1+\tau}  \left( k +2 p\cdot rS \right)\,, \\
   C^\mathrm{L}  &=  0. \label{eq:ABC_lepton3}
\end{align}
To evaluate $\mathrm{L}_\mathrm{t}$, the lepton spin asymmetry with the spin vector in the scattering plane and perpendicular to the lepton momentum, we substitute $\left( p \cdot r S \right) = 0 $ and $\left( k \cdot r S \right) = 2 r \sqrt{ \tau \nu^2 -  (1+\tau)  (\tau + r^2)^2 }/\sqrt{\left( \nu + r^2 - \tau \right)^2-4r^2}$ in Eqs.~(\ref{eq:ABC_lepton1})$-$(\ref{eq:ABC_lepton3}) above. To evaluate $\mathrm{L}_\mathrm{l}$, the lepton spin asymmetry with the spin vector in the scattering plane and parallel to the lepton momentum, we substitute $2 \left( p \cdot r S \right) = \sqrt{\left( \nu + r^2 - \tau \right)^2-4r^2}$ and $ \left( k \cdot r S \right) =  - \left(\left(r^2 - \tau \right) \nu + \left( \tau + r^2 \right)^2\right)/\sqrt{\left( \nu + r^2 - \tau \right)^2-4r^2}$ in Eqs.~(\ref{eq:ABC_lepton1})$-$(\ref{eq:ABC_lepton3}) above. The transverse lepton single-spin asymmetry $\mathrm{L_t}$ vanishes at forward and backward angles. Up to lepton-mass-suppressed terms, the longitudinal single-spin asymmetry reaches its extremum reflecting the chiral nature of the weak interaction, i.e., $\mathrm{L_l} = -\eta + \mathrm{O} \left( m_\ell^2 \right)$.

Spin polarization asymmetries provide a novel probe of nucleon structure that is complementary to unpolarized cross section measurements. In contrast to a typical polarization experiment in strong and electromagnetic interactions, spin asymmetries in weak interactions are large. In a polarization experiment, flux normalization errors and detector systematics largely cancel in the asymmetry expression paving the way to clean probes of the nucleon axial and pseudoscalar form factors from polarization observables.

\section{Polarization observables with muon and electron neutrinos}
\label{sec3}

In this Section, we evaluate polarization observables in charged current quasielastic neutrino-nucleon scattering with muon and electron neutrinos. We provide the unpolarized cross section and single-spin asymmetries for muon neutrino beam of hundred MeV energies when the pseudoscalar form factor can sizably contribute to spin-dependent observables.

\begin{figure}[H]
          \centering
          \includegraphics[height=0.44\textwidth]{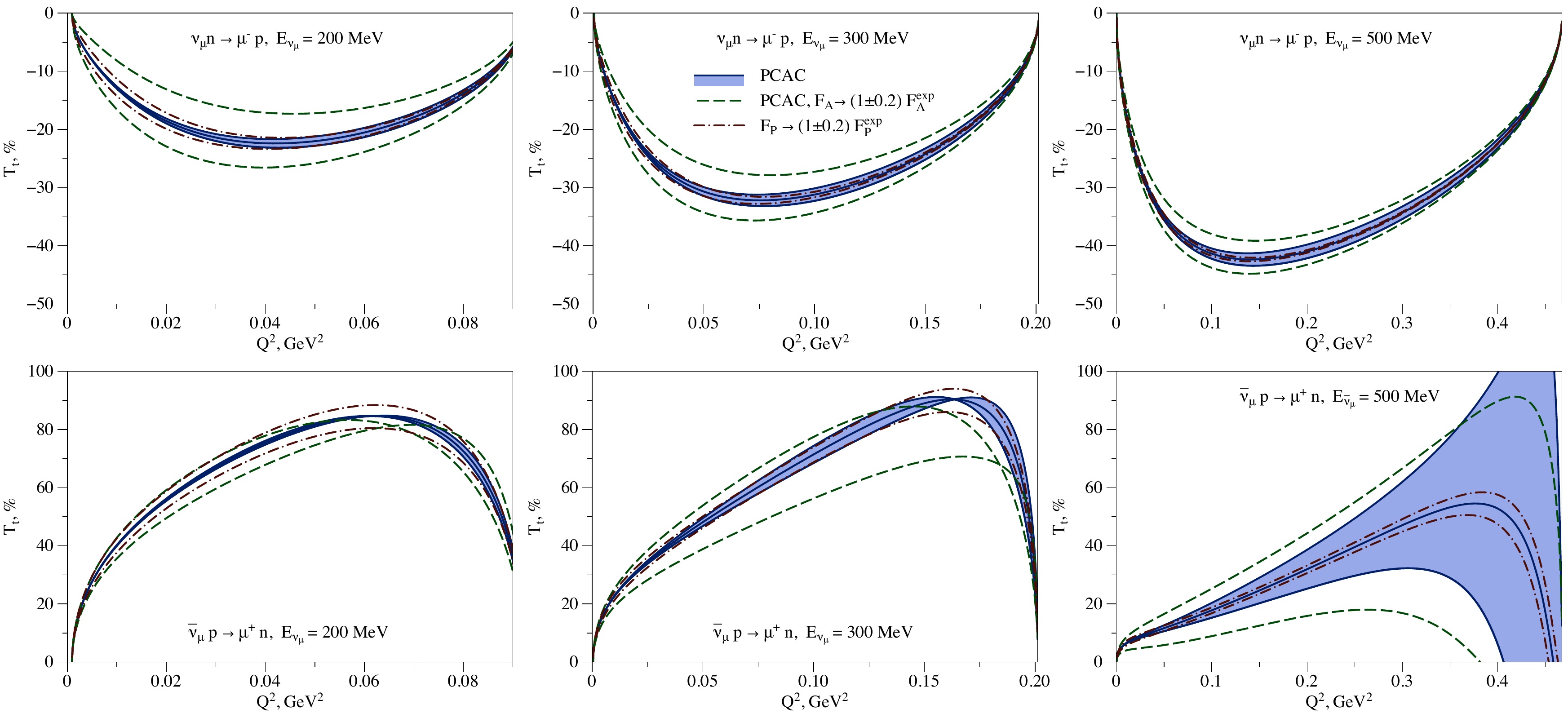}              
          \caption{The spin asymmetry $\mathrm{T}_\mathrm{t}$ in charged current quasielastic muon-neutrino-neutron (upper panel) and antineutrino-proton (lower panel) scattering at neutrino beam energies $E_\nu = 200~\mathrm{MeV},~300~\mathrm{MeV}$, and $500~\mathrm{MeV}$. Errors on all plots (blue band labeled as PCAC) are propagated from the fit parameters and covariance matrices of~\cite{Meyer:2016oeg,Borah:2020gte} and added in quadrature for the axial and electromagnetic form factors. Green dashed lines correspond to observables with the axial form factor changed by 20~\% from the expected value while keeping the PCAC ansatz for the pseudoscalar form factor. Red dash-dotted lines represent a 20~\% variation of the pseudoscalar form factor from the expected value.
    \label{fig:Tt_asymmetries}}
\end{figure}

\begin{figure}[H]
          \centering
          \includegraphics[height=0.44\textwidth]{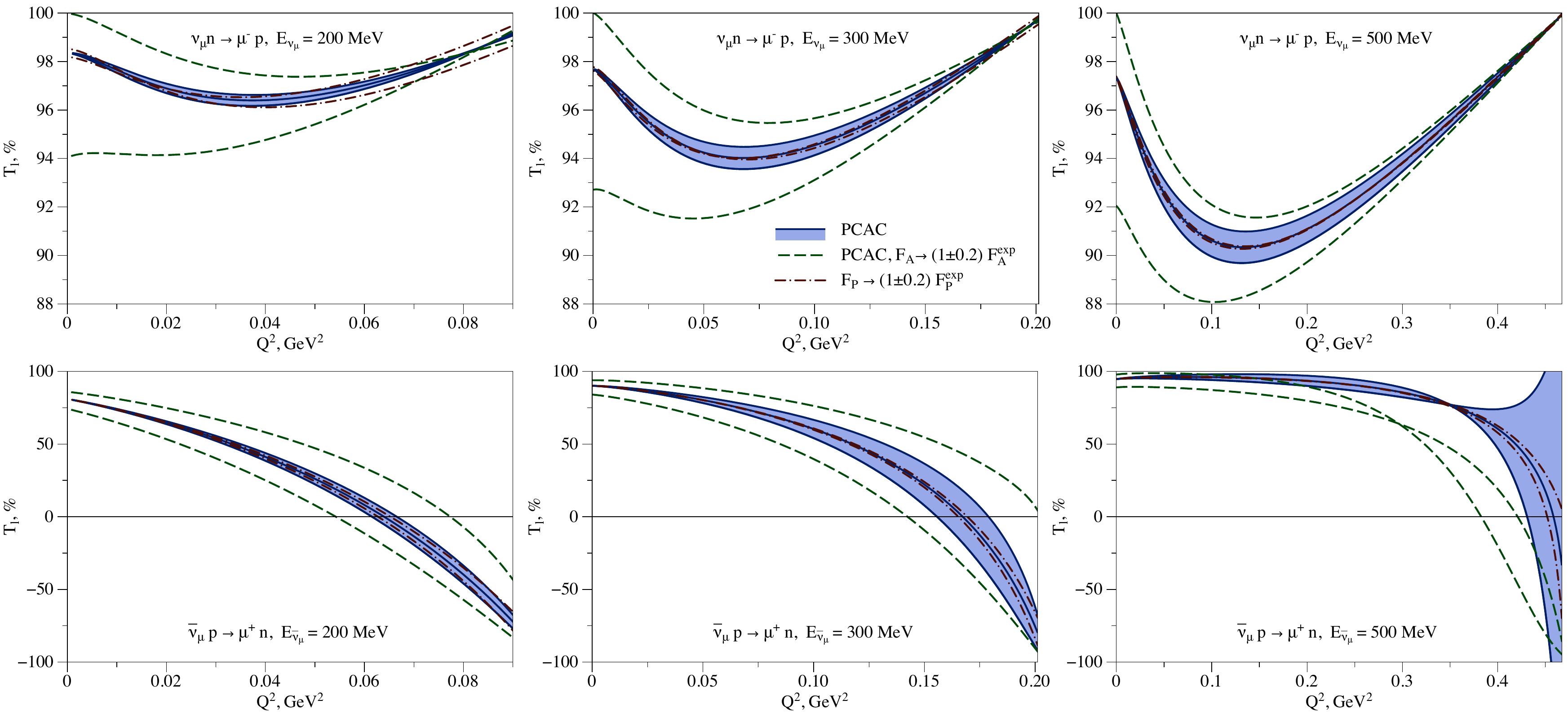}              
          \caption{The spin asymmetry $\mathrm{T}_\mathrm{l}$ in charged current quasielastic muon-neutrino-neutron (upper panel) and antineutrino-proton (lower panel) scattering at neutrino beam energies $E_\nu = 200~\mathrm{MeV},~300~\mathrm{MeV}$, and $500~\mathrm{MeV}$.
    \label{fig:Tl_asymmetries}}
\end{figure}

\begin{figure}[H]
          \centering
          \includegraphics[height=0.44\textwidth]{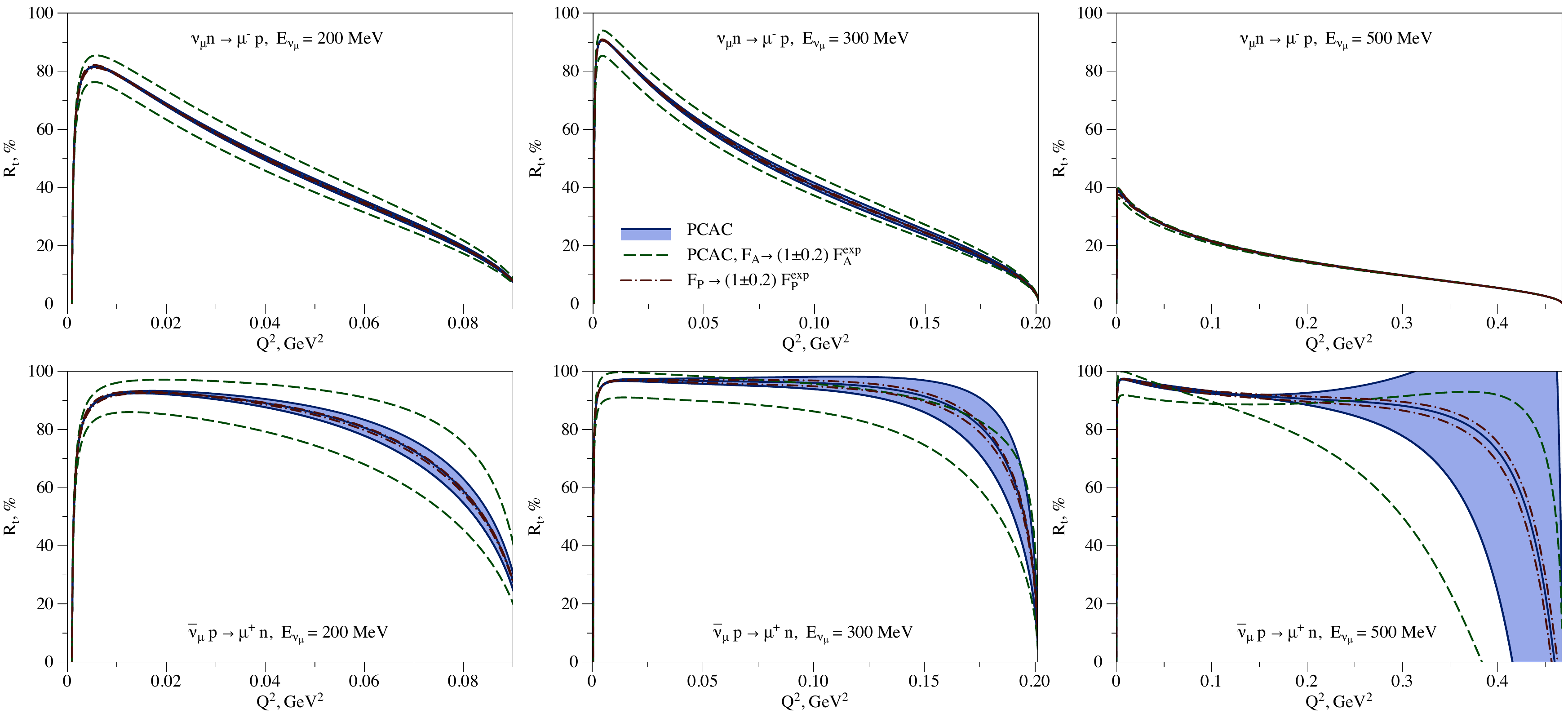}              
          \caption{The spin asymmetry $\mathrm{R}_\mathrm{t}$ in charged current quasielastic muon neutrino-neutron (upper panel) and antineutrino-proton (lower panel) scattering at neutrino beam energies $E_\nu = 200~\mathrm{MeV},~300~\mathrm{MeV}$, and $500~\mathrm{MeV}$.
    \label{fig:Rt_asymmetries}}
\end{figure}

\begin{figure}[H]
          \centering
          \includegraphics[height=0.44\textwidth]{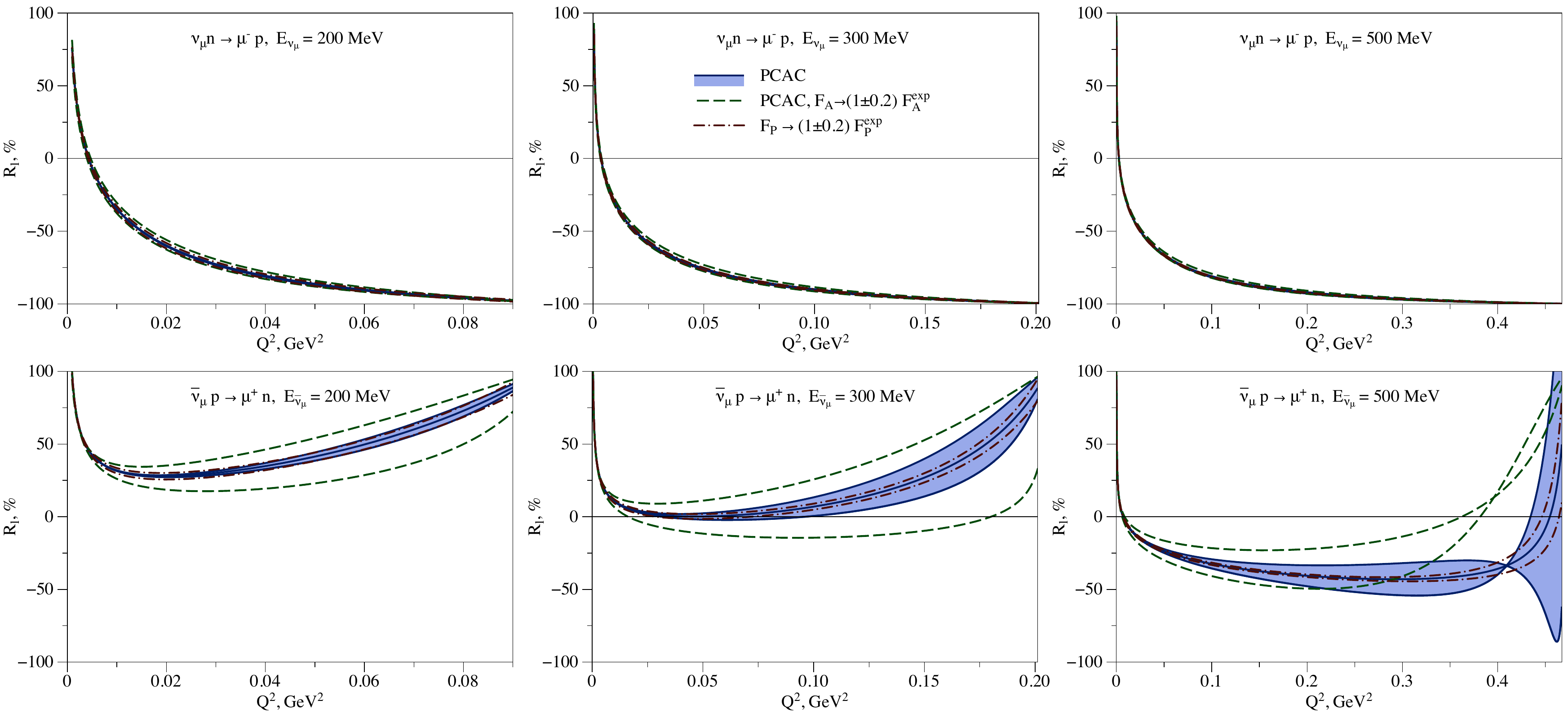}              
          \caption{The spin asymmetry $\mathrm{R}_\mathrm{l}$ in charged current quasielastic muon-neutrino-neutron (upper panel) and antineutrino-proton (lower panel) scattering at neutrino beam energies $E_\nu = 200~\mathrm{MeV},~300~\mathrm{MeV}$, and $500~\mathrm{MeV}$.
    \label{fig:Rl_asymmetries}}
\end{figure}

\begin{figure}[H]
          \centering
          \includegraphics[height=0.44\textwidth]{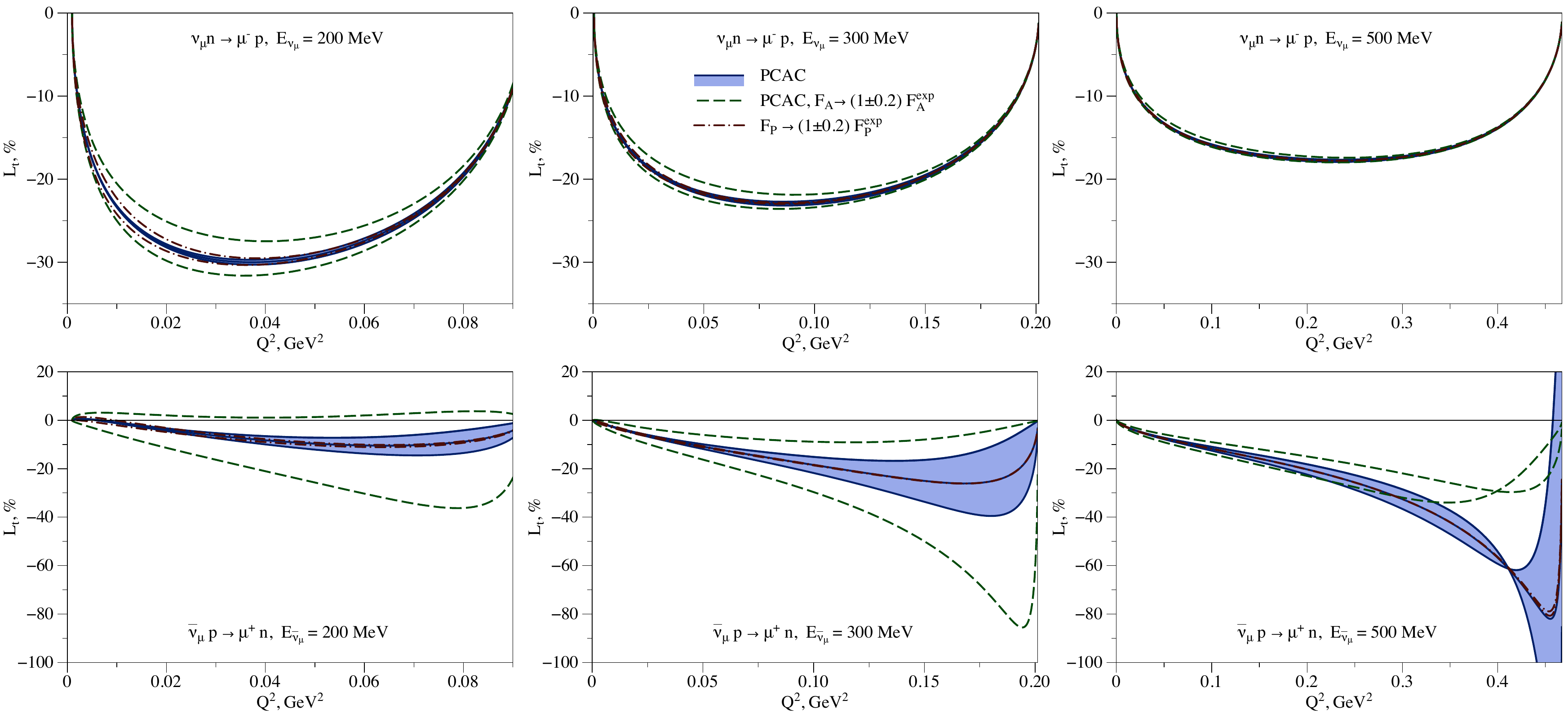}              
          \caption{The spin asymmetry $\mathrm{L}_\mathrm{t}$ in charged current quasielastic muon-neutrino-neutron (upper panel) and antineutrino-proton (lower panel) scattering at neutrino beam energies $E_\nu = 200~\mathrm{MeV},~300~\mathrm{MeV}$, and $500~\mathrm{MeV}$.
    \label{fig:Lt_asymmetries}}
\end{figure}

\begin{figure}[H]
          \centering
          \includegraphics[height=0.44\textwidth]{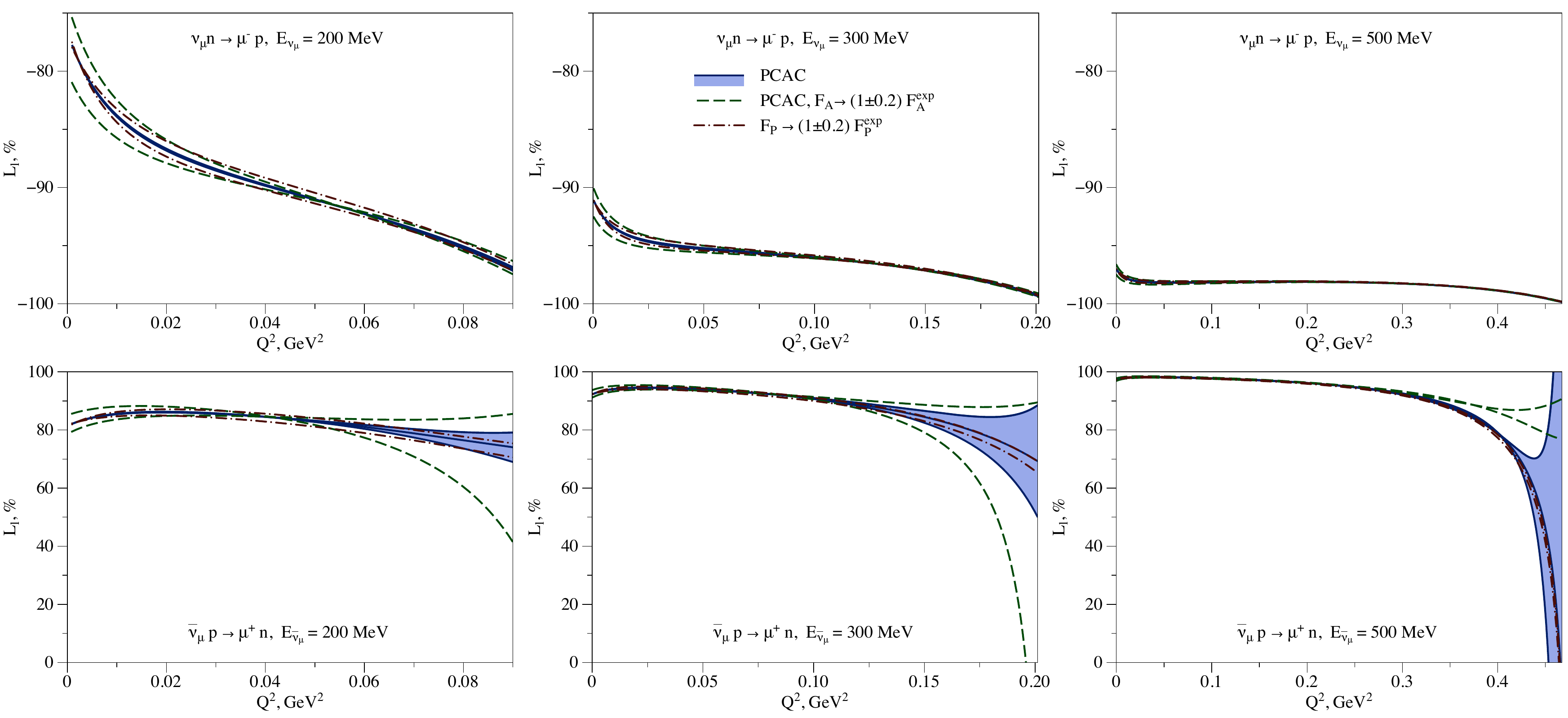}              
          \caption{The spin asymmetry $\mathrm{L}_\mathrm{l}$ in charged current quasielastic muon-neutrino-neutron (upper panel) and antineutrino-proton (lower panel) scattering at neutrino beam energies $E_\nu = 200~\mathrm{MeV},~300~\mathrm{MeV}$, and $500~\mathrm{MeV}$.
    \label{fig:Ll_asymmetries}}
\end{figure}

The pseudoscalar form factor contribution in the scattering of $\nu_e$ and $\bar{\nu}_e$ is suppressed by factors $m^2_e/E^2_\nu,~m^2_e/M^2$, and $m^2_e/\left(M E_\nu\right)$, and is therefore negligible at energies of accelerator experiments. The pseudoscalar form factor contribution in the scattering of $\nu_\mu$ and $\bar{\nu}_\mu$ is negligible at neutrino beam energies $E_\nu$ above the nucleon mass $E_\nu \gtrsim M$. At lower neutrino beam energies, around a few hundred MeV, the pseudoscalar form factor becomes reachable by making use of polarization observables. This influence persists down to the muon production threshold; however, in this limit total event rates become very small; beam energy of roughly 150$-$250 MeV is therefore ideally suited to maximizing the sensitivity to the pseudoscalar form factor in scattering experiments with muon (anti)neutrinos. In Figs.~\ref{fig:Tt_asymmetries}$-$\ref{fig:Ll_asymmetries}, we present all nonvanishing single-spin asymmetries in muon (anti)neutrino scattering at above-threshold energies when the pseudoscalar contribution can be sizable. For illustration, we substitute nucleon form factors from~\cite{Meyer:2016oeg,Borah:2020gte} assuming partial conservation of the axial-vector current and pion-pole dominance (PCAC ansatz) for the pseudoscalar form factor: $ F_P(Q^2) =2M^2/ \left( m_\pi^2 + Q^2\right) F_A(Q^2)$ (though PCAC ansatz can be valid only at $Q^2 \lesssim \Lambda^2_\mathrm{QCD} $). We propagate errors for the axial and electromagnetic form factors separately and add the uncertainties in quadrature. We also compare central values varying the axial form factor by 20~\% versus varying the pseudoscalar form factor from PCAC value by 20~\%.\footnote{The normalization of axial and pseudoscalar form factors are known pretty well from neutron decay and muon capture rates on hydrogen, so our variations can represent deviations only away from $Q^2 = 0$.} According to definitions above, all asymmetries are in the range $[ -100, 100 ]~\%$. 

The target transverse single-spin asymmetry, Fig.~\ref{fig:Tt_asymmetries}, is negative in neutrino-neutron scattering and positive in antineutrino-proton scattering. In neutrino-neutron CCQE, the asymmetry $\mathrm{T_t}$ decreases with the neutrino beam energy down to an absolute value of 20-40~\% level. This asymmetry is more sensitive to the axial than to the pseudoscalar form factor. However, the change in this asymmetry to the central value after the variation of $F_A$ by 20~\% is typically below 5~\%. In antineutrino-proton scattering, such a change can exceed 10$-$20~\% at neutrino beam energies $300~\mathrm{MeV}$, and $500~\mathrm{MeV}$. The target transverse single-spin asymmetry $\mathrm{T_t}$ can reach up to $85~\%$ in antineutrino-proton scattering. For all energies, neutrino and antineutrino scattering, the asymmetry $\mathrm{T_t}$ as a function of momentum transfer first raises from 0 to maximum and then decreases to 0 as pointed in Sec.~\ref{sec2}. Assuming the PCAC ansatz, the asymmetry is predicted up to a few percent level besides antineutrino-proton scattering at antineutrino beam energies of order $500~\mathrm{MeV}$ when the error reaches $10-30~\%$ level. The target longitudinal single-spin asymmetry, Fig.~\ref{fig:Tl_asymmetries}, is known at the percent level for neutrino-neutron scattering and at 10-20~\% level for antineutrino-proton scattering. As for the target transverse asymmetry, this asymmetry is more sensitive to the axial form factor than to the pseudoscalar form factor. The asymmetry $\mathrm{T_t}$ in antineutrino-proton scattering is sensitive also to the pseudoscalar form factor at lowest neutrino energies. The asymmetry $\mathrm{T_l}$ is close to 100~\% in ${\nu}_\mu n \to \mu^- p$ and varies almost over all allowed range in $\bar{\nu}_\mu p \to \mu^+ n$. As for the asymmetry $\mathrm{T_t}$, antineutrino-proton scattering is more promising for studies of the axial nucleon structure.

The recoil nucleon single-spin asymmetries, see Figs.~\ref{fig:Rt_asymmetries} and~\ref{fig:Rl_asymmetries}, are also more sensitive to the axial than to the pseudoscalar form factor. Transverse recoil asymmetry is positive over all kinematic ranges of low-energy neutrinos. $\mathrm{R_t}$ is known better for neutrino-neutron than for antineutrino-proton scattering. This asymmetry does not show sensitivity to the pseudoscalar form factor and can be exploited for extractions of the axial form factor in the case of antineutrino-proton CCQE when variations of $\mathrm{R_t}$ reach $10-50~\%$ level changing the axial form factor by 20~\%. The longitudinal recoil single-spin asymmetry can be negative or positive for low-energy kinematics. As for the transverse asymmetry, $\mathrm{R_l}$ is known better in neutrino-neutron scattering assuming PCAC ansatz and pion-pole dominance, while antineutrino-proton scattering can be used for complementary extractions of the axial form factor. Moreover, the asymmetry $\mathrm{R_l}$ shows sensitivity to the pseudoscalar form factor at lowest muon neutrino energies; see the left lower panel in Fig.~\ref{fig:Rl_asymmetries}.

The transverse recoil lepton single-spin asymmetry $\mathrm{L_t}$, see Fig.~\ref{fig:Lt_asymmetries}, is sensitive mainly to the axial form factor at low-energy kinematic region. The asymmetry $\mathrm{L_t}$ is negative both in neutrino-neutron and antineutrino-proton scattering. In both cases, the transverse recoil lepton single-spin asymmetry does not exceed $30~\%$ level by an absolute value besides antineutrino-proton scattering with the highest beam energy in Fig.~\ref{fig:Lt_asymmetries}. By an absolute value, the asymmetry $\mathrm{L_t}$ has an inverted U-shaped behavior as a function of the momentum transfer. As for recoil and target asymmetries described above, the PCAC-based prediction is more precise for neutrino-neutron scattering while the antineutrino-proton scattering is more promising for constraints of the axial structure. The longitudinal recoil lepton asymmetry $\mathrm{L_l}$, see Fig.~\ref{fig:Ll_asymmetries}, is negative in neutrino-neutron and positive in antineutrino-proton CCQE scattering. This asymmetry is typically above $80~\%$ by an absolute value. It approaches the maximum absolute value increasing the energy of the neutrino beam. In both ${\nu}_\mu n \to \mu^- p$ and $\bar{\nu}_\mu p \to \mu^+ n$, this asymmetry is predicted with percent or even subpercent level of precision which makes the extraction of the nucleon axial form factor from $\mathrm{L_l}$ a challenging task. At a certain kinematic region at low energies, the single-spin polarization observable $\mathrm{L_l}$ in antineutrino-proton scattering is more sensitive to the pseudoscalar than to the axial form factor, see the left lower panel in Fig.~\ref{fig:Ll_asymmetries}.
\begin{figure}[H]
          \centering
          \includegraphics[height=0.22\textwidth]{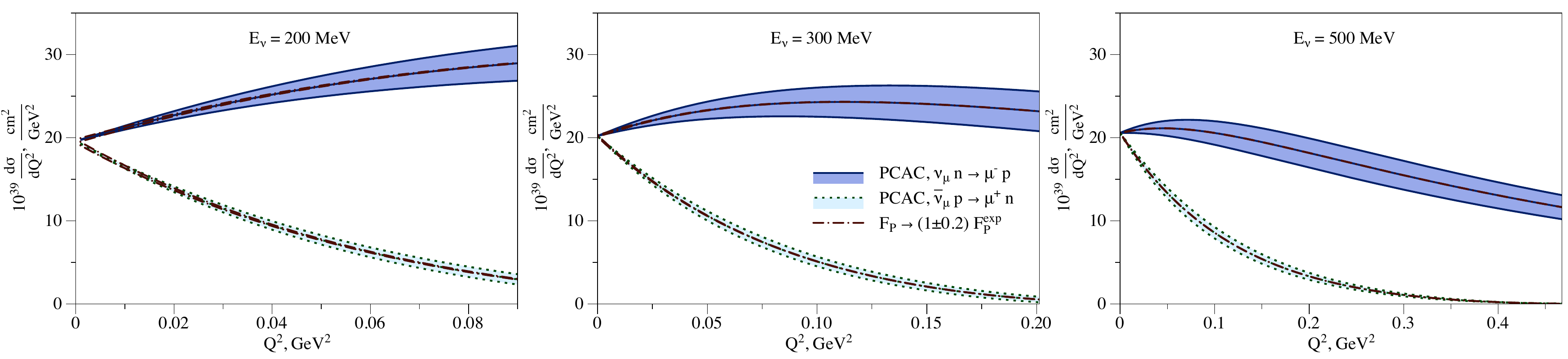}              
          \caption{Unpolarized cross sections in charged current quasielastic muon-neutrino-neutron and antineutrino-proton scattering at neutrino beam energies $E_\nu = 200~\mathrm{MeV},~300~\mathrm{MeV}$, and $500~\mathrm{MeV}$.
    \label{fig:cross_sections}}
\end{figure}

Nevertheless both the axial and the pseudoscalar form factors are constrained by the chiral perturbation theory at low~$Q^2$~\cite{Bernard:1998gv,Fuchs:2002zz,Kaiser:2003dr,Schindler:2006it,Lutz:2020dfi}, it is remarkable that both the pseudoscalar and the axial form factors can be measured simultaneously in experiments with the muon (anti)neutrino beam of a few hundred MeV energy and polarized particles, though axial contributions to asymmetries are slightly kinematically enhanced compared to pseudoscalar ones. To have an idea of how large event rates can be, we present the unpolarized cross sections both for neutrino-neutron and antineutrino-proton processes in Fig.~\ref{fig:cross_sections}. The neutrino-neutron unpolarized cross section as a function of the momentum transfer increases at low values of energy and momentum transfer and falls with $Q^2$ at higher values while the antineutrino-proton cross sections always decrease with momentum transfer. As one can notice from Fig.~\ref{fig:cross_sections}, the unpolarized cross section for the muon flavor at hundred MeV energies is not sensitive to the pseudoscalar form factor even changing the latter by 20~\%. CCQE cross sections are enhanced in the low-$Q^2$ region which can give potentially larger event rates for all asymmetries. Having larger cross section, neutrino-neutron scattering would provide more events for a given flux and number of nucleons compared to antineutrino-proton scattering. However, total event rates are suppressed by phase space volume compared to neutrinos of GeV energies resulting in smaller total cross sections.

\section{Axial form factor at GeV energies}
\label{sec4}

In this Section, we calculate polarization observables at most common energies of accelerator neutrinos and present the most promising single-spin asymmetries. These asymmetries can give us a complementary way for the measurements of the nucleon axial form factor. We provide also asymmetries averaged over the typical neutrino flux at modern neutrino oscillation experiments.

\begin{figure}[H]
          \centering
          \includegraphics[height=0.22\textwidth]{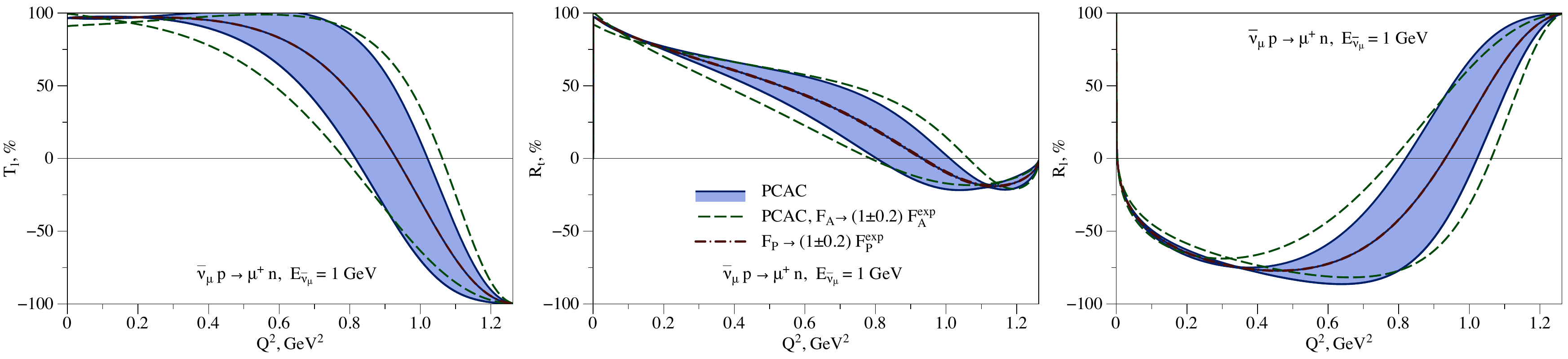}              
          \caption{The spin asymmetries $\mathrm{T}_\mathrm{l}, \mathrm{R}_\mathrm{t}, \mathrm{R}_\mathrm{l}$ in charged current quasielastic muon-antineutrino-proton scattering at antineutrino beam energy $E_{\bar{\nu}_\mu} = 1~\mathrm{GeV}$. Asymmetries for electron (anti)neutrino scattering are indistinguishable from results on these figures.
    \label{fig:asymmetries_enhanced_at_GeV}}
\end{figure}

Almost all asymmetries at 1~GeV energy and above require a few percent or subpercent precision to contribute significantly to the global uncertainties on the axial structure. Only $\mathrm{T}_\mathrm{l}, \mathrm{R}_\mathrm{t}, \mathrm{R}_\mathrm{l}$ in $\bar{\nu}_\ell p \to \ell^+ n$ at GeV energies and above are of practical interest. We present these observables in Fig.~\ref{fig:asymmetries_enhanced_at_GeV}. These asymmetries change as a function of the momentum transfer over a wide range of allowed values and are sensitive mainly to the axial but not to the pseudoscalar form factor. Similar values of error bands and dashed lines representing a 20~\% variation of the axial form factor in Fig.~\ref{fig:asymmetries_enhanced_at_GeV} confirm that the axial form factor is the main source of uncertainties predicting single-spin asymmetries. Other asymmetries in antineutrino-proton and neutrino-neutron scattering are either too small to be measured by the first polarization experiments or are not sensitive enough to the form factor $F_A$. 

\begin{figure}[H]
          \centering
          \includegraphics[height=0.22\textwidth]{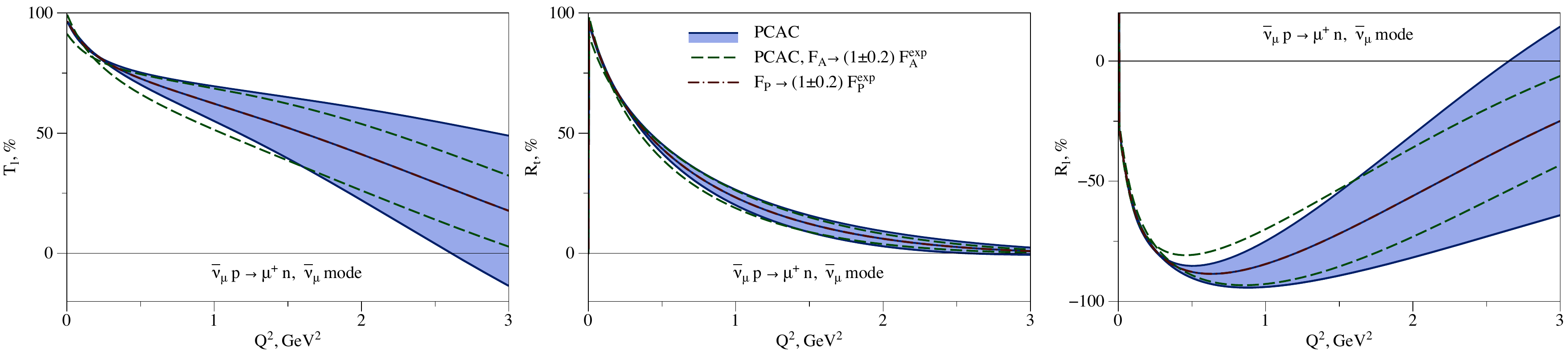}              
          \caption{The spin asymmetries $\mathrm{T}_\mathrm{l}, \mathrm{R}_\mathrm{t}, \mathrm{R}_\mathrm{l}$ in charged current quasielastic antineutrino-proton scattering averaged over expected DUNE near-detector flux.
    \label{fig:asymmetries_averaged_over_DUNE}}
\end{figure}

High-intensity fluxes of modern and future accelerator experiments provide an attractive opportunity for precise measurements of percent level effects. To select the most promising experimental observables as would be relevant for a spin-polarized target installed along the DUNE beamline as proposed in \cite{Snowmass2021LoInuHD}, we average over the anticipated flux profiles of the DUNE Near Detector~\cite{Alion:2016uaj,dune_page} at Fermilab. Neglecting detector details, we present a closer to experiment result in Fig.~\ref{fig:asymmetries_averaged_over_DUNE}. Adding high-energy flux components, the asymmetry $\mathrm{R}_\mathrm{t}$ loses sensitivity to the axial structure. However, $\mathrm{T}_\mathrm{l}$ and $\mathrm{R}_\mathrm{t}$ provide a complementary to the unpolarized cross section probe of the axial form factor. 

\begin{figure}[H]
          \centering
          \includegraphics[height=0.22\textwidth]{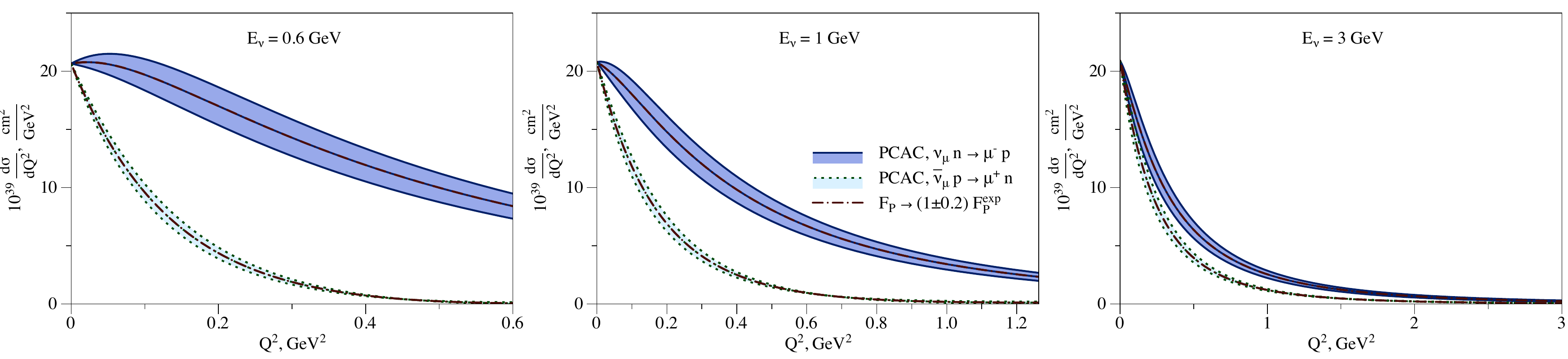}              
          \caption{Unpolarized cross sections in charged current quasielastic muon-neutrino-neutron and antineutrino-proton scattering at neutrino beam energies $E_{\bar{\nu}_\mu} = 0.6~\mathrm{GeV},~1~\mathrm{GeV}$, and $3~\mathrm{GeV}$.
    \label{fig:cross_sections_at_GeV}}
\end{figure}

To get an idea on possible event rates, we provide the unpolarized cross section for a few energies in Fig.~\ref{fig:cross_sections_at_GeV}. At lower energies and momentum transfers, neutrino-neutron CCQE cross sections largely exceed antineutrino-proton cross sections while both saturate at higher energies. 

\begin{figure}[H]
          \centering
          \includegraphics[height=0.22\textwidth]{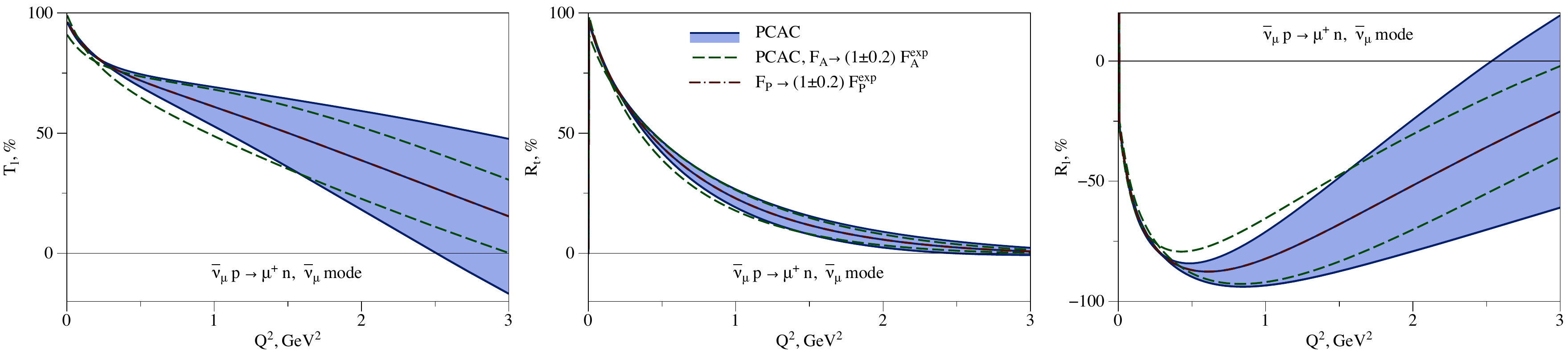}              
          \caption{Same as Fig.~\ref{fig:asymmetries_averaged_over_DUNE} but for the averaging over the T2K flux.
    \label{fig:asymmetries_averaged_over_T2K}}
\end{figure}

As an illustrative example of a neutrino flux that peaks at lower energies, we imagine a spin-polarized facility with T2K near detector flux and average over the typical T2K flux profile~\cite{Abe:2012av,Abe:2015awa} as an input. We provide flux-averaged asymmetries in Fig.~\ref{fig:asymmetries_averaged_over_T2K}. Much like for DUNE flux, for T2K flux asymmetries $\mathrm{T}_\mathrm{l}$ and $\mathrm{R}_\mathrm{t}$ in antineutrino-proton scattering are the most interesting for studying the nucleon axial structure. 

\section{Polarization observables with tau neutrinos}
\label{sec5}

In this Section, we evaluate polarization observables in charged current quasielastic neutrino-nucleon scattering with tau neutrinos. We study the sensitivity of the unpolarized cross section and single-spin asymmetries to the axial and pseudoscalar form factors.

\begin{figure}[H]
          \centering
          \includegraphics[height=0.22\textwidth]{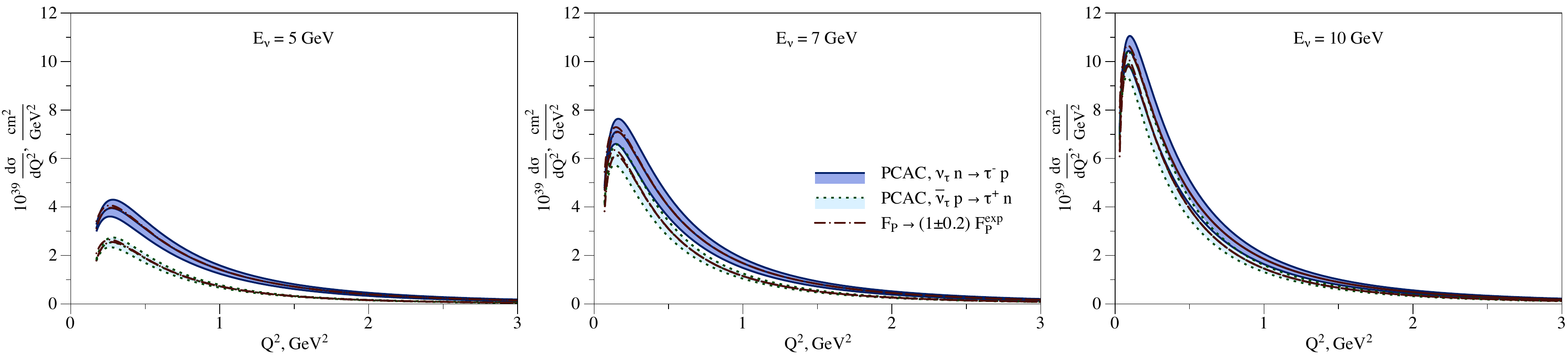}              
          \caption{Unpolarized cross sections in charged current quasielastic tau-neutrino-neutron and antineutrino-proton scattering at neutrino beam energies $E_\nu = 5~\mathrm{GeV},~7~\mathrm{GeV}$, and $10~\mathrm{GeV}$.
    \label{fig:cross_sections2}}
\end{figure}

\begin{figure}[H]
          \centering
          \includegraphics[height=0.44\textwidth]{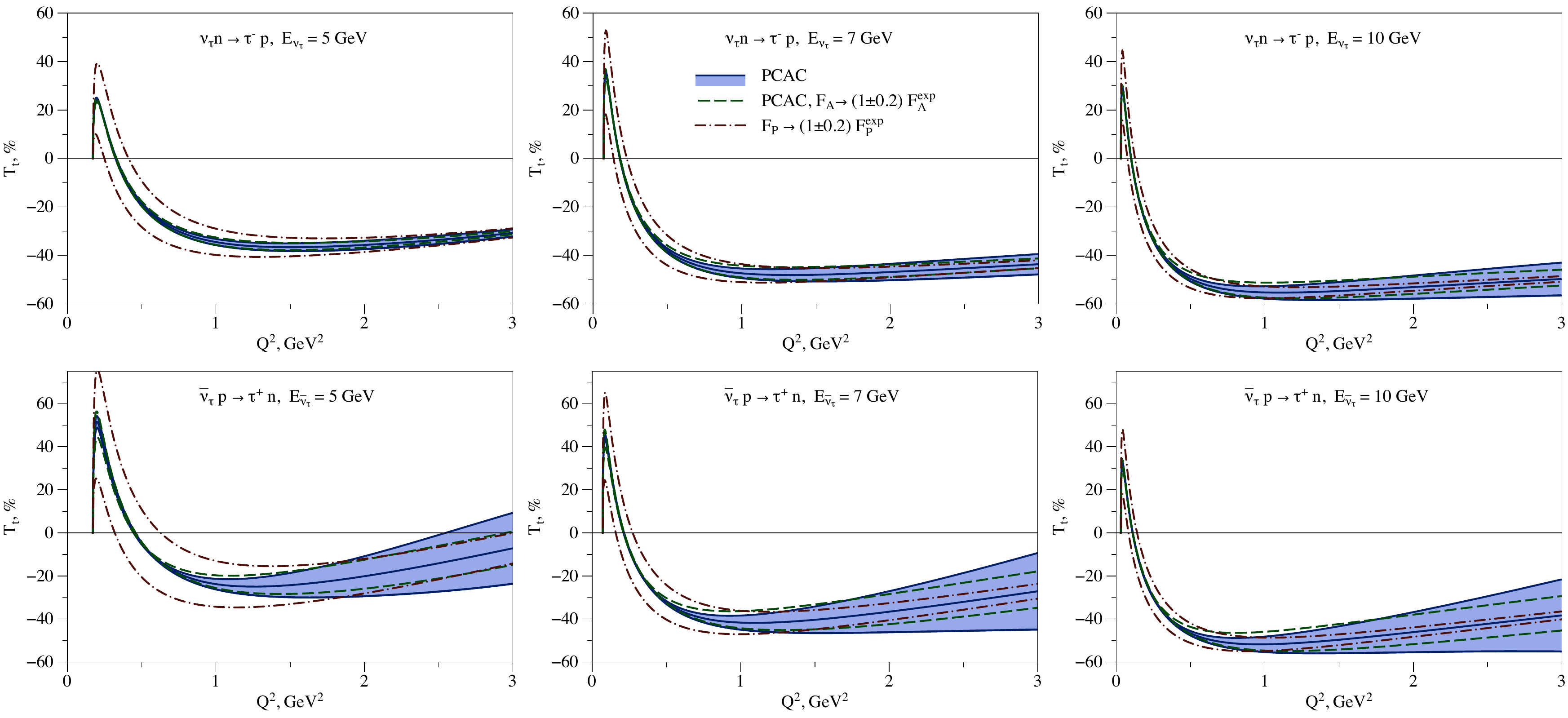}              
          \caption{The spin asymmetry $\mathrm{T}_\mathrm{t}$ in charged current quasielastic tau-neutrino-neutron (upper panel) and antineutrino-proton (lower panel) scattering at neutrino beam energies $E_\nu = 5~\mathrm{GeV},~7~\mathrm{GeV}$, and $10~\mathrm{GeV}$.
    \label{fig:Tt_asymmetries2}}
\end{figure}

\begin{figure}[H]
          \centering
          \includegraphics[height=0.44\textwidth]{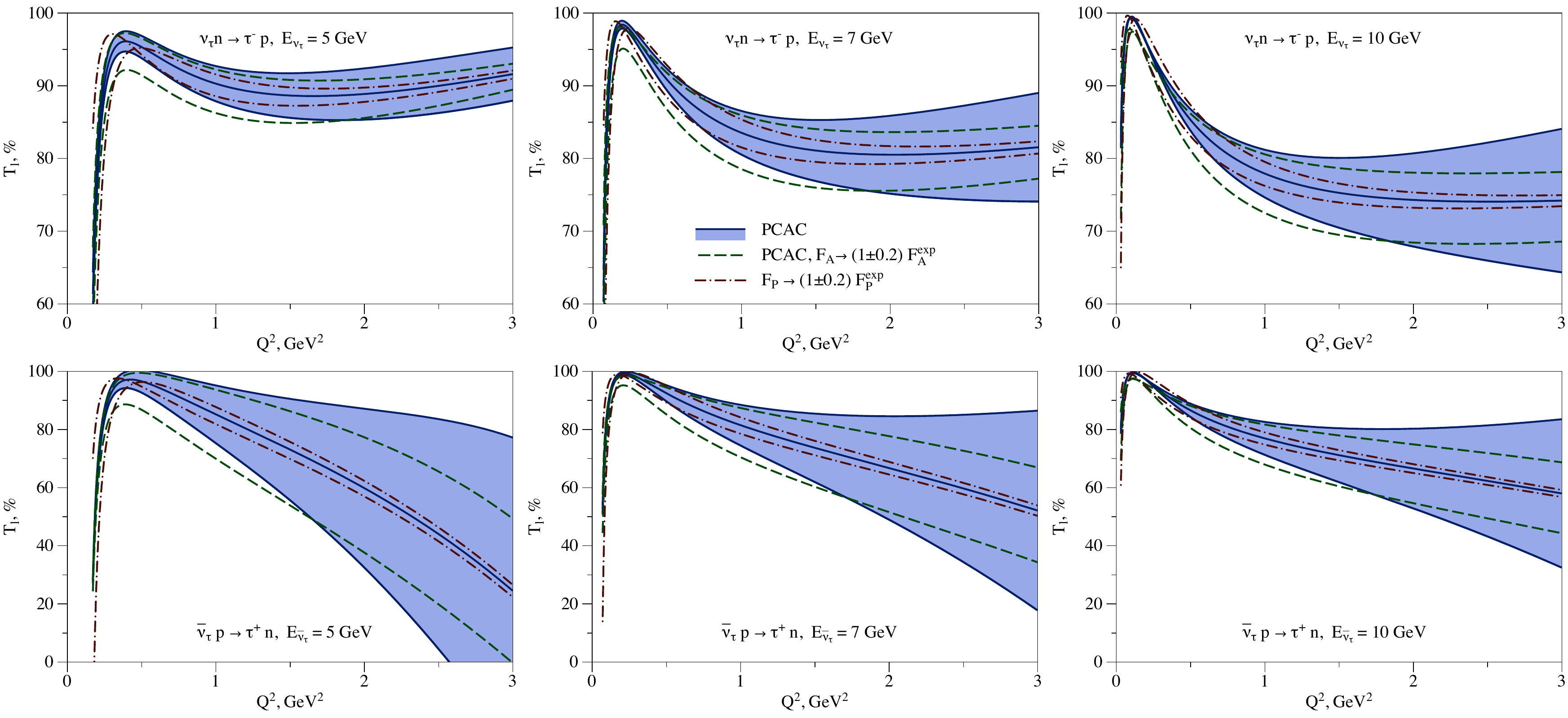}              
          \caption{The spin asymmetry $\mathrm{T}_\mathrm{l}$ in charged current quasielastic tau-neutrino-neutron (upper panel) and antineutrino-proton (lower panel) scattering at neutrino beam energies $E_\nu = 5~\mathrm{GeV},~7~\mathrm{GeV}$, and $10~\mathrm{GeV}$.
    \label{fig:Tl_asymmetries2}}
\end{figure}

\begin{figure}[H]
          \centering
          \includegraphics[height=0.44\textwidth]{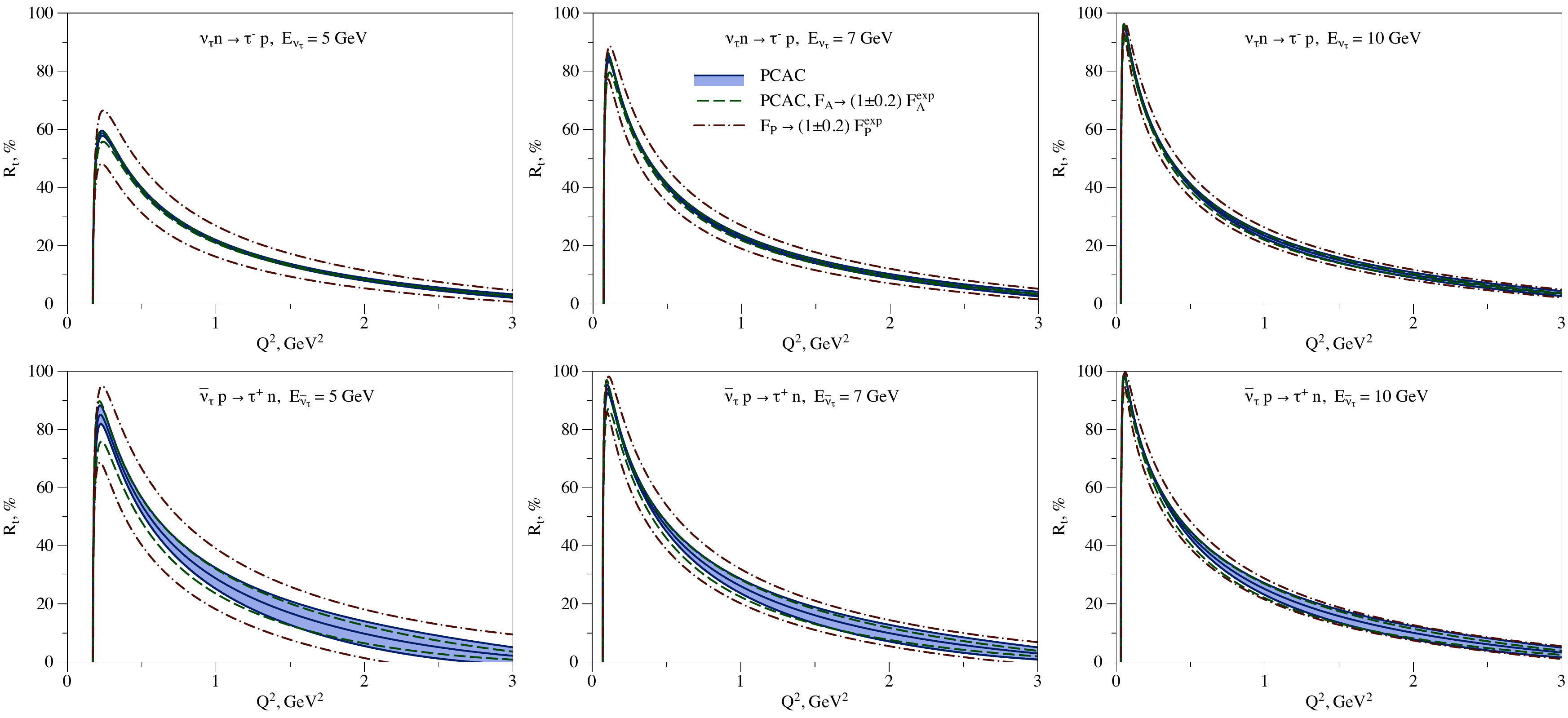}              
          \caption{The spin asymmetry $\mathrm{R}_\mathrm{t}$ in charged current quasielastic tau-neutrino-neutron (upper panel) and antineutrino-proton (lower panel) scattering at neutrino beam energies $E_\nu = 5~\mathrm{GeV},~7~\mathrm{GeV}$, and $10~\mathrm{GeV}$.
    \label{fig:Rt_asymmetries2}}
\end{figure}

\begin{figure}[H]
          \centering
          \includegraphics[height=0.44\textwidth]{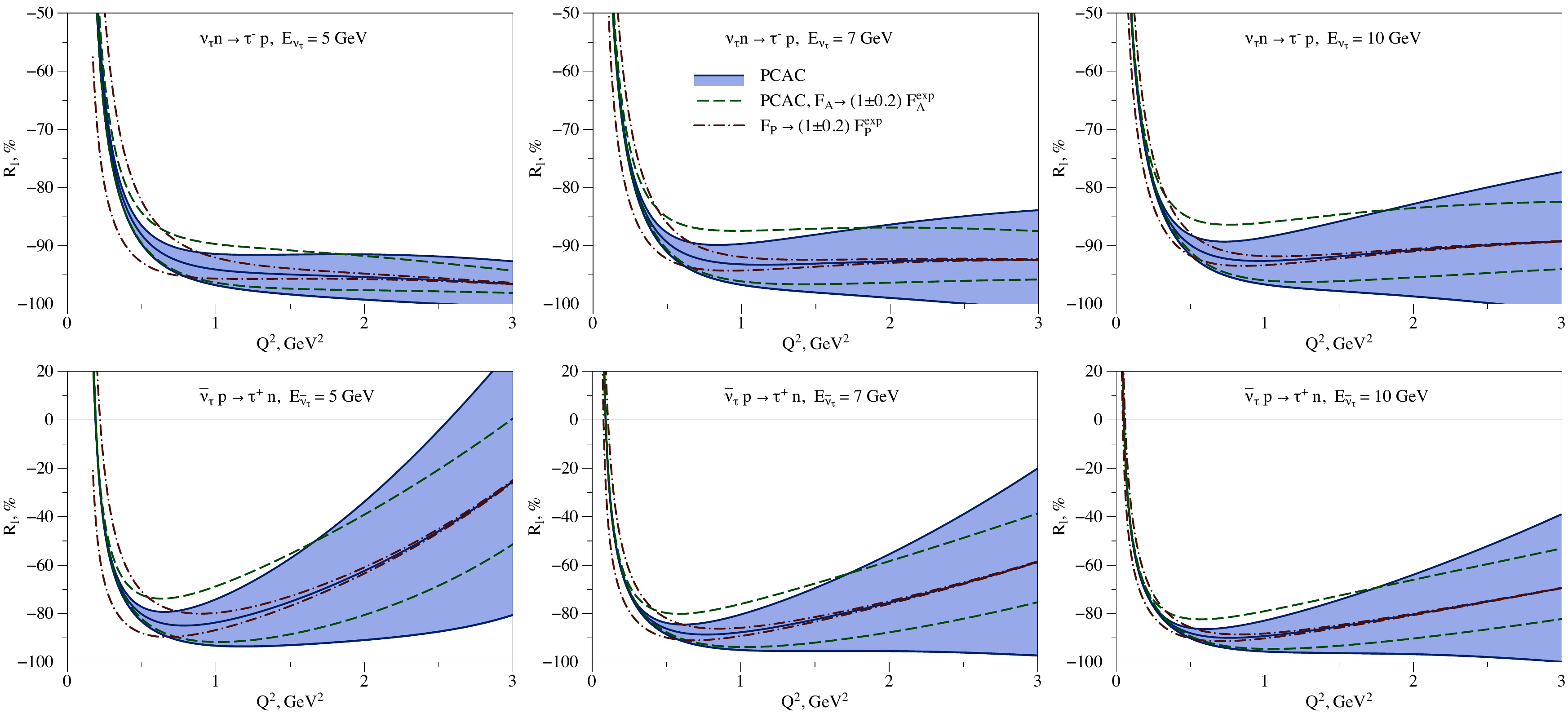}              
          \caption{The spin asymmetry $\mathrm{R}_\mathrm{l}$ in charged current quasielastic tau-neutrino-neutron (upper panel) and antineutrino-proton (lower panel) scattering at neutrino beam energies $E_\nu = 5~\mathrm{GeV},~7~\mathrm{GeV}$, and $10~\mathrm{GeV}$.
    \label{fig:Rl_asymmetries2}}
\end{figure}

\begin{figure}[H]
          \centering
          \includegraphics[height=0.44\textwidth]{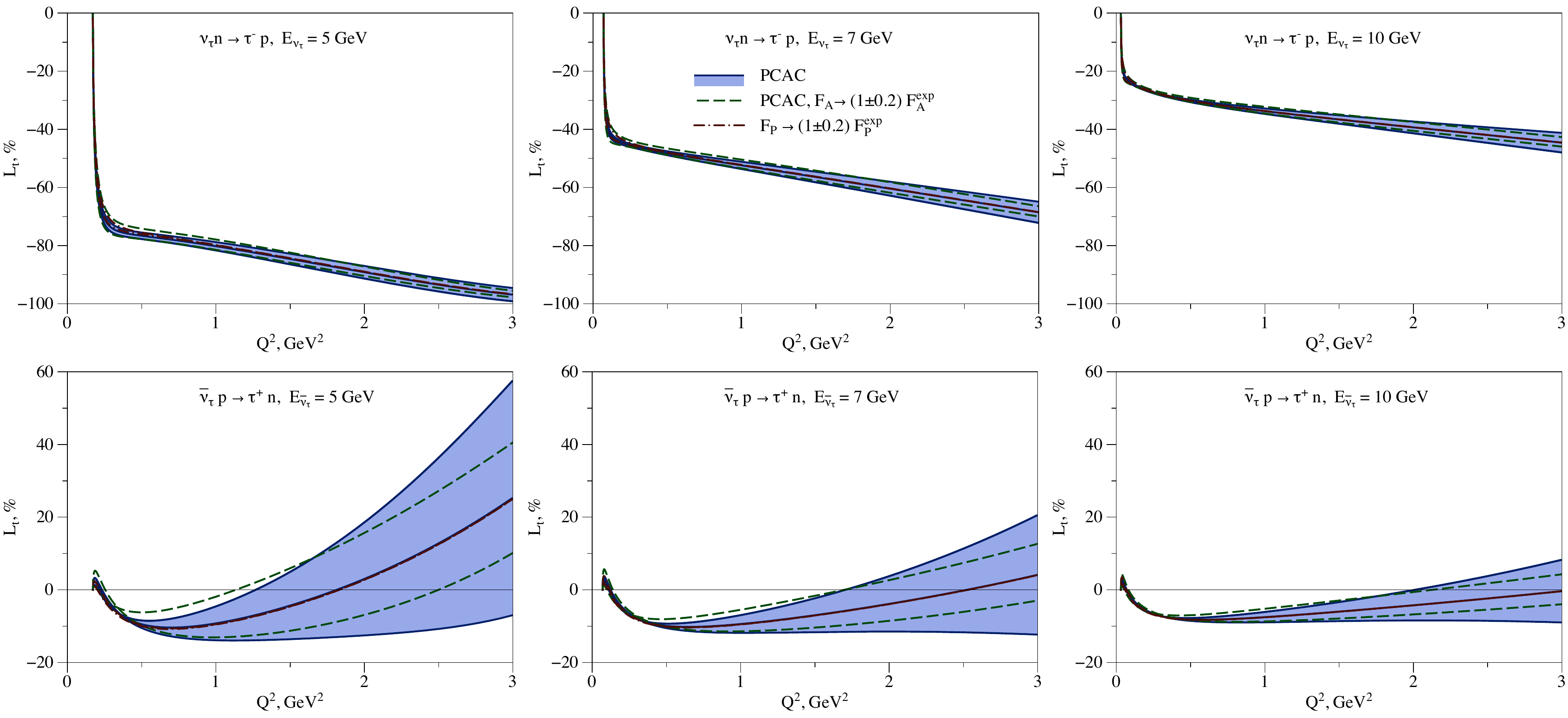}              
          \caption{The spin asymmetry $\mathrm{L}_\mathrm{t}$ in charged current quasielastic tau-neutrino-neutron (upper panel) and antineutrino-proton (lower panel) scattering at neutrino beam energies $E_\nu = 5~\mathrm{GeV},~7~\mathrm{GeV}$, and $10~\mathrm{GeV}$.
    \label{fig:Lt_asymmetries2}}
\end{figure}

\begin{figure}[H]
          \centering
          \includegraphics[height=0.44\textwidth]{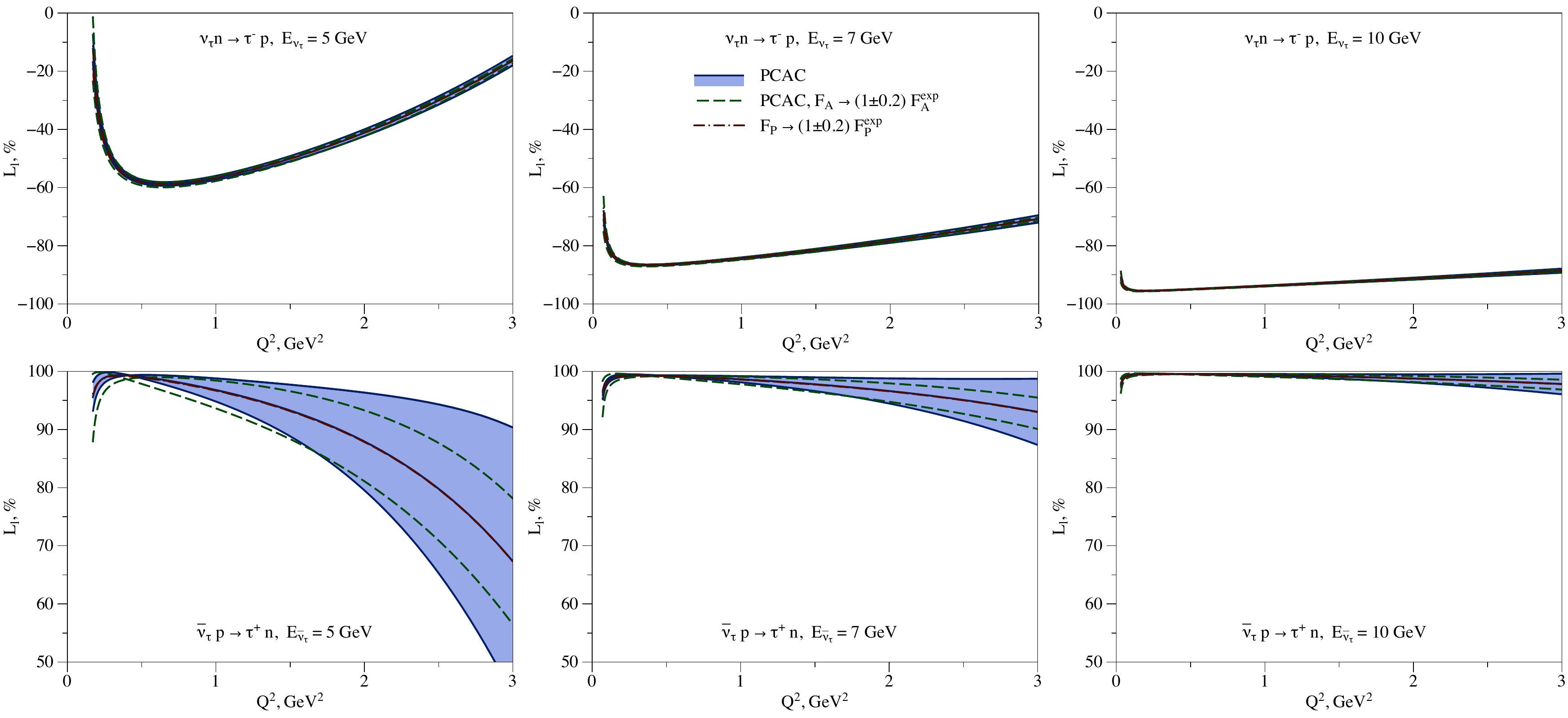}              
          \caption{The spin asymmetry $\mathrm{L}_\mathrm{l}$ in charged current quasielastic tau-neutrino-neutron (upper panel) and antineutrino-proton (lower panel) scattering at neutrino beam energies $E_\nu = 5~\mathrm{GeV},~7~\mathrm{GeV}$, and $10~\mathrm{GeV}$.
    \label{fig:Ll_asymmetries2}}
\end{figure}

The contribution of the pseudoscalar form factor to CCQE observables with tau (anti)neutrinos is not suppressed by a small lepton mass factor making beams of tau neutrinos attractive above the tau production threshold, $E_\nu \gtrsim 3.5~\mathrm{GeV}$. Contrary to the unpolarized cross sections shown in Fig.~\ref{fig:cross_sections2} that has some sensitivity to the pseudoscalar form factor at lowest momentum transfers, spin asymmetries in scattering of tau (anti)neutrinos are very sensitive to the pseudoscalar form factor, see Figs.~\ref{fig:Tt_asymmetries2}$-$\ref{fig:Ll_asymmetries2} for details. The energy range on these figures corresponds to the range of modern accelerator-based neutrino beams above the tau-production threshold. Above the tau-production threshold, the unpolarized cross section increases with the neutrino beam energy. Near the threshold, the neutrino-neutron unpolarized cross section is above the antineutrino-proton cross section. This difference vanishes increasing the neutrino beam energies and becomes small within the uncertainty at tau neutrino beam energy $8-10~\mathrm{GeV}$ when the cross section is almost saturated.

The transverse target single-spin asymmetry, see Fig.~\ref{fig:Tt_asymmetries2}, shows similar behavior in neutrino-neutron and antineutrino-proton scattering. In both cases, the asymmetry $\mathrm{T_t}$ as a function of the momentum transfer rapidly increases from $0$  up to a maximum value and then decreases to $-(40-60)~\%$ [$-(10-40)~\%$ at lower beam energies] at $Q^2 \simeq 1~\mathrm{GeV}^2$ and above where the asymmetry is almost constant up to $Q^2 \simeq 3~\mathrm{GeV}^2$. At $Q^2 \lesssim 1~\mathrm{GeV}^2$, the transverse target asymmetry is more sensitive to the pseudoscalar than to the axial form factor and vise verse at higher momentum transfers. Contrary, the longitudinal target single-spin asymmetry shown in Fig.~\ref{fig:Tl_asymmetries2} is more sensitive to the axial than to the pseudoscalar form factor over the whole kinematic range $Q^2 \lesssim 3~\mathrm{GeV}^2$. The asymmetry $\mathrm{T_l}$ is positive and typically above $60~\%$. It has the largest values at lower neutrino beam energies. For neutrino-neutron and antineutrino-proton CCQE, the asymmetry $\mathrm{T_l}$ shows similar dependence on the momentum transfer and energy at $Q^2 \lesssim 1~\mathrm{GeV}^2$. Above this momentum transfer, the longitudinal target asymmetry $\mathrm{T_l}$ in neutrino-nucleon scattering is more flat than in the antineutrino-proton case.

The transverse recoil single-spin asymmetry, see Fig.~\ref{fig:Rt_asymmetries2}, is more sensitive to the pseudoscalar than to the axial form factor up to $Q^2 \simeq 3~\mathrm{GeV}^2$ both in ${\nu}_\mu n \to \mu^- p$ and $\bar{\nu}_\mu p \to \mu^+ n$ processes. The asymmetry $\mathrm{R_t}$ is positive and has similar behavior in these two reactions. Assuming the PCAC ansatz and pion-pole dominance, $\mathrm{R_t}$ is predicted better in neutrino-neutron scattering. As the transverse target single-spin asymmetry, the transverse recoil single-spin asymmetry as a function of the momentum transfer rapidly increases from 0 to the maximum value and then slowly decreases down to 5$-$10~\% level at $Q^2 \simeq 3~\mathrm{GeV}^2$. The longitudinal recoil single-spin asymmetry, see Fig.~\ref{fig:Rl_asymmetries2}, is mainly negative besides the region of very small momentum transfers. Its absolute value increases up to $Q^2 \simeq 0.5-1~\mathrm{GeV}^2$ and saturates at $90-95~\%$ in neutrino-neutron scattering while slowly decreases in antineutrino-proton scattering. The asymmetry $\mathrm{R_l}$ is sensitive more to the pseudoscalar than to the axial form factor at lower $Q^2 \lesssim 0.5~\mathrm{GeV}^2$ and rapidly loses this sensitivity at larger values of the momentum transfer. As for many other asymmetries and unpolarized cross section, the longitudinal recoil single-spin asymmetry is predicted with smaller uncertainties for ${\nu}_\mu n \to \mu^- p$ than for $\bar{\nu}_\mu p \to \mu^+ n$ process. This asymmetry does not show big deviations varying the neutrino beam energy.

The tau polarization impacts the angular distribution of the tau decay products and can be reconstructed from the kinematics of daughter particles. That is why the recoil tau asymmetry has attracted a lot of attention~\cite{Hagiwara:2003di,Hagiwara:2004gs,Graczyk:2004vg,Graczyk:2004uy,Bourrely:2004iy,Kuzmin:2004ke,Kuzmin:2004yb,Aoki:2005wb,Aoki:2005kc,Sobczyk:2019urm,Fatima:2020pvv}. Both transverse and longitudinal recoil tau asymmetries shown in Figs.~\ref{fig:Lt_asymmetries2} and \ref{fig:Ll_asymmetries2} are sensitive mainly to the axial form factor. Variations of the axial form factor by $20~\%$ change the recoil lepton single-spin asymmetries in neutrino-neutron scattering at the percent or even subpercent level. The transverse recoil lepton asymmetry in antineutrino-proton CCQE does not exceed 20$-$30~\% by magnitude. This asymmetry in neutrino-neutron scattering and the longitudinal recoil tau single-spin asymmetry in ${\nu}_\mu n \to \mu^- p$ are negative while the asymmetry $\mathrm{L_l}$ in $\bar{\nu}_\mu p \to \mu^+ n$ process is positive. The asymmetry $\mathrm{L_t}$ in antineutrino-proton scattering can be either positive or negative. The asymmetry $\mathrm{L_t}$ in neutrino-neutron scattering decreases by an absolute value increasing the neutrino beam energy while the asymmetry $\mathrm{L_l}$ in neutrino-neutron scattering increases by an absolute value. The asymmetry $\mathrm{L_l}$ in antineutrino-proton scattering is typically above the level $90-95~\%$.

The recoil and target asymmetries at low $Q^2$ are more sensitive to the pseudoscalar than to the axial form factor. The lepton polarization asymmetries $\mathrm{L}_\mathrm{l}$ and $\mathrm{L}_\mathrm{t}$ are sensitive only to the axial form factor. A sizable dataset with $\nu_\tau, \bar{\nu}_\tau$ could allow us to access the pseudoscalar form factor from neutrino scattering data. Future investigations of anticipated data at DUNE far detector~\cite{Acciarri:2015uup,Abi:2020evt}, SHIP facility~\cite{Anelli:2015pba,Alekhin:2015byh}, and DsTau experiment~\cite{Aoki:2019jry} at CERN accounting for the corresponding nuclear physics effects would be of great interest.

\section{Conclusions}
\label{sec6}

In conclusion, we have studied the sensitivity to axial nucleon structure of single-spin asymmetries in (anti)neutrino charged current quasielastic scattering on free nucleons. Many of these asymmetries provide much better sensitivity to the pseudoscalar form factor compared to the unpolarized cross section. The pseudoscalar form factor can be extracted either from asymmetries in the scattering cross sections of muon (anti)neutrinos at hundreds of MeV energy performing very precise experiments or from transverse target and recoil nucleon asymmetries in the scattering cross sections of tau (anti)neutrinos above the tau production threshold $E_\nu \gtrsim 3.5~\mathrm{GeV}$. The axial form factor can be extracted from polarization observables at GeV energies in a complementary way from recoil longitudinal $\mathrm{R}_\mathrm{l}$ and target longitudinal $\mathrm{T}_\mathrm{l}$ asymmetries in $\bar{\nu}_e p \to e^+ n$ and $\bar{\nu}_\mu p \to \mu^+ n$. The first measurement of polarization observables in neutrino-nucleon scattering experiments could provide a new test of the Standard Model of particle physics, complementary information on the axial form factor, and an independent way to measure the pseudoscalar form factor.

\section*{Acknowledgments}

We thank Peter Filip, Richard Hill, and Adam Aurisano for useful discussions, Tom Junk and Ryan Plestid for numerous useful suggestions regarding the text and presentation. This work was supported by the U.S. Department of Energy, Office of Science, Office of High Energy Physics, under Award Number DE-SC0019095. 
Fermilab is operated by Fermi Research Alliance, LLC under Contract No. DE-AC02-07CH11359 with the United States Department of Energy. The author would like to acknowledge the Fermilab theory group and the theory group of Institute for Nuclear Physics at Johannes Gutenberg-Universit\"at Mainz for warm hospitality and support. The work of O.T. is supported by the Visiting Scholars Award Program of the Universities Research Association. FeynCalc~\cite{Mertig:1990an,Shtabovenko:2016sxi} and Mathematica~\cite{Mathematica} were useful in this work.

\appendix

\bibliography{polarization}{}

\end{document}